%% file: NeurIPS_main.tex
\documentclass{article}

\PassOptionsToPackage{numbers, compress}{natbib}
\usepackage[final]{neurips_2020}
\usepackage[utf8]{inputenc} 
\usepackage[T1]{fontenc}    
\usepackage{hyperref}       
\usepackage{url}            
\usepackage{booktabs}       
\usepackage{amsfonts}       
\usepackage{nicefrac}       
\usepackage{microtype}      
\usepackage{algorithm}
\usepackage{algorithmic}
\usepackage{mathtools}
\usepackage{enumitem}

\usepackage{bm}

\usepackage{graphicx}
\usepackage{float}
\usepackage{xcolor}

\title{FinRL-Meta: Market Environments and Benchmarks for Data-Driven Financial Reinforcement Learning}

\author{%
  Xiao-Yang Liu$^{1*}$, Ziyi Xia$^1\thanks{Equal contribution.}$ , Jingyang Rui$^2$, Jiechao Gao$^3$, Hongyang Yang$^1$, \\ \textbf{Ming Zhu$^4$, Christina Dan Wang$^{5\dagger}$, Zhaoran Wang$^6$, Jian Guo$^7 $\thanks{Corresponding authors.}}\\
  $^1$Columbia University; $^2$The University of Hongkong; $^3$University of Virginia; \\$^4$SIAT CAS; $^5$New York University (Shanghai); $^6$Northwestern University; \\$^7$IDEA Research, International Digital Economy Academy\\
  \texttt{\{xl2427, zx2325, hy2500\}@columbia.edu, christina.wang@nyu.edu,} \\
  \texttt{zhaoranwang@northwestern.edu, guojian@idea.edu.cn}\\
}

\begin{document}

\maketitle

\begin{abstract}
  Finance is a particularly difficult playground for deep reinforcement learning. However, establishing high-quality market environments and benchmarks for financial reinforcement learning is challenging due to three major factors, namely, low signal-to-noise ratio of financial data, survivorship bias of historical data, and model overfitting in the backtesting stage. In this paper, we present an openly accessible FinRL-Meta library that has been actively maintained by the AI4Finance community. First, following a DataOps paradigm, we will provide hundreds of market environments through an automatic pipeline that collects dynamic datasets from real-world markets and processes them into gym-style market environments. Second, we reproduce popular papers as stepping stones for users to design new trading strategies. We also deploy the library on cloud platforms so that users can visualize their own results and assess the relative performance via community-wise competitions. Third, FinRL-Meta provides tens of Jupyter/Python demos organized into a curriculum and a documentation website to serve the rapidly growing community. FinRL-Meta is available at: \url{https://github.com/AI4Finance-Foundation/FinRL-Meta} 
\end{abstract}

\input{NeurIPS_2020_version/Section1_Introduction}

\input{NeurIPS_2020_version/Section2_ExistingWorks}

\input{NeurIPS_2020_version/Section3_Architecture}

\input{NeurIPS_2020_version/Section3_Environment}

\input{NeurIPS_2020_version/Section4_Evaluation}

\section{Conclusion}

In this paper, we obeyed the DataOps paradigm and developed a FinRL-Meta library that provides openly accessible dynamic financial datasets and reproducible benchmarks. For future work, FinRL-Meta aims to build a universe of financial market environments, like the XLand environment \cite{team2021open}. To improve the performance for the large-scale markets, we are exploiting GPU-based massive parallel simulation such as Isaac Gym \cite{makoviychuk2021isaac}. Moreover, it will be interesting to explore the evolutionary perspectives \cite{gupta2021embodied,scholl2021market,finrl_podracer_2021, liu2021podracer} to simulate the markets. We believe that FinRL-Meta will provide insights into complex market phenomena and offer guidance for financial regulations.

\section*{Acknowledgement}

We thank Mr. Tao Liu (IDEA Research, International Digital Economy Academy) for technical support of computing platform on this research project. Ming Zhu was supported by National Natural Science Foundations of China (Grant No. 61902387). Christina Dan Wang is supported in part by National Natural Science Foundation of China (NNSFC) grant 11901395 and Shanghai Pujiang Program, China 19PJ1408200.

\newpage

\medskip
\small
\bibliographystyle{plain}
\bibliography{ref}

\input{NeurIPS_2020_version/appendix}

\end{document}

%% file: NeurIPS_2020_version/Section1_Introduction.tex
\section{Introduction}

Finance is a particularly challenging playground for deep reinforcement learning (DRL) \cite{sutton2018reinforcement,hambly2021recent}, including investigating market fragility \cite{raberto2001agent}, developing profitable strategies \cite{xiong2018practical, yang2020deep,zhang2020deep}, and assessing portfolio risk \cite{lussange2021modelling, bao2019multiagent}. However, establishing near-real market environments and benchmarks on financial reinforcement learning are challenging due to three major factors, namely, low signal-to-noise ratio (SNR) of financial data, survivorship bias of historical data, and model overfitting in the backtesting stage. Such a \textit{simulation-to-reality gap} \cite{DulacArnold2020AnEI,dulac2019challenges} degrades the performance of DRL strategies in real markets. Therefore, high-quality market environments and DRL benchmarks are crucial for the research and industrialization of data-driven financial reinforcement learning.

Existing works have applied various DRL algorithms in financial applications \cite{lussange2021modelling, liu2021finrl, karpe2020multi, pricope2021deep}. Many of them have shown better trading performance in terms of cumulative return and Sharpe ratio. Several recent works \cite{lussange2021modelling, amrouni2021abides, karpe2020multi} showed the great potential of DRL-based market simulators that are not publicly available yet. Therefore, these works are difficult to reproduce. The FinRL library \cite{liu2020finrl,liu2021finrl} provided an open-source framework for financial reinforcement learning. However, it focused on guaranteeing reproducibility of backtesting performance while several market environments were provided.  A workshop version of FinRL-Meta \cite{finrl_meta_2021} provided data processors to access and clean unstructured market data, but it did not provide benchmarks back then.

\begin{figure}[t]
\centering
\includegraphics[scale = 0.082]{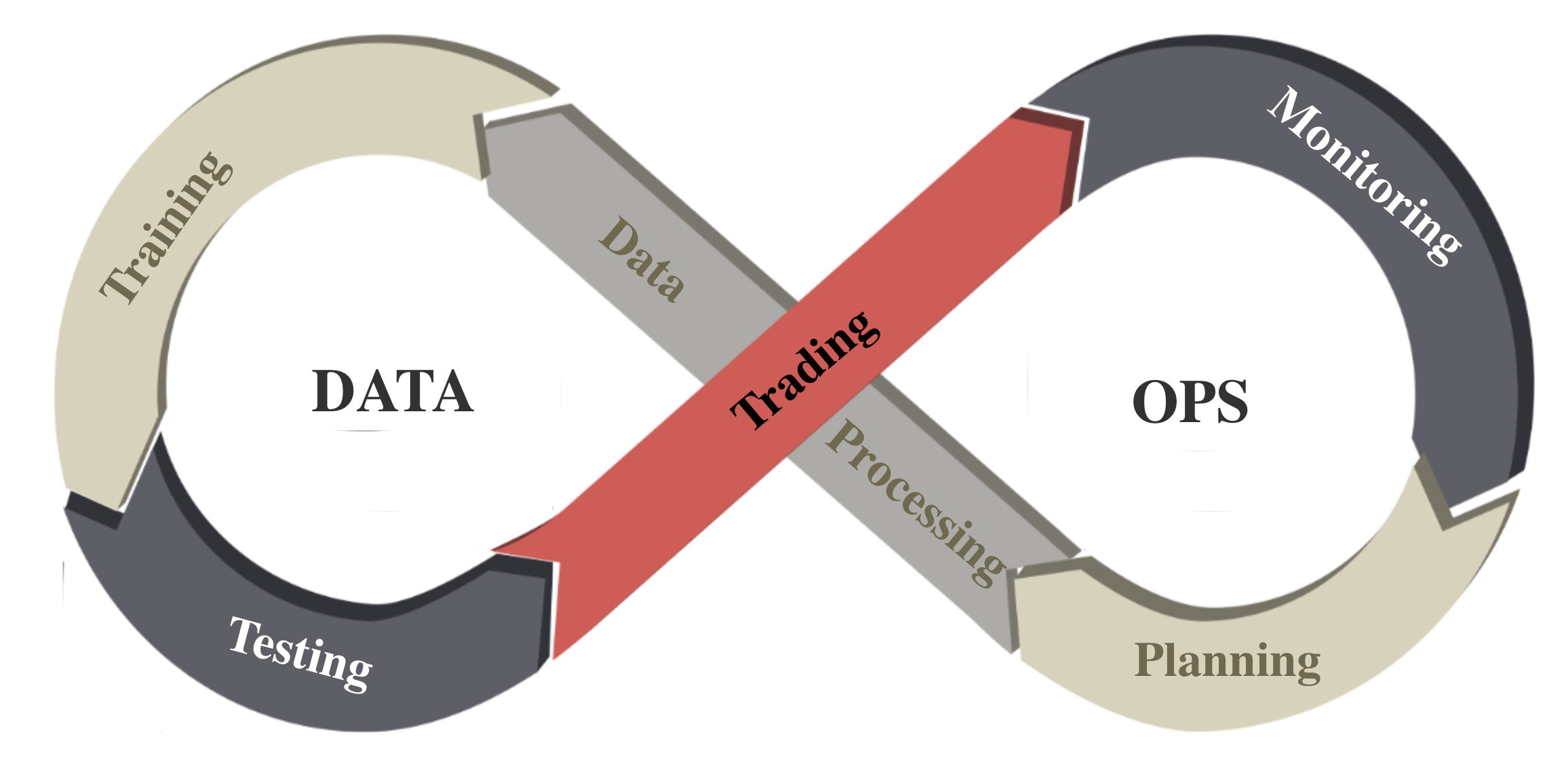}
\hspace{5mm}
\includegraphics[scale = 0.17]{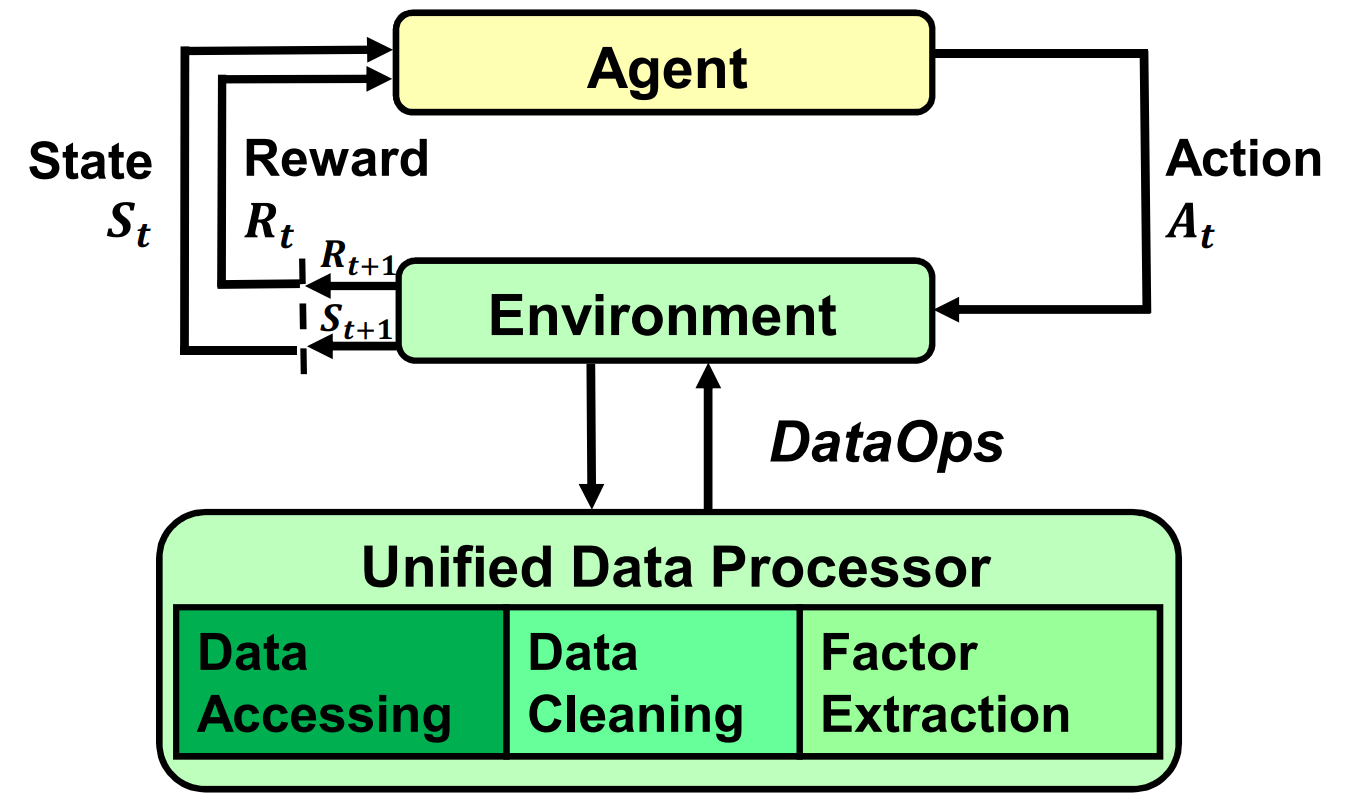}
\caption{DataOps paradigm (left) and data-driven financial reinforcement learning (right).}
\label{fig:conventional vs neofinrl}
\vspace{-4mm}
\end{figure}

The \textsf{DataOps} paradigm \cite{DataOps, atwal2019practical, ereth2018dataops} refers to a set of practices, processes, and technologies that combines automated data engineering and agile development \cite{ereth2018dataops}. It helps reduce the cycle time of data engineering and improve data quality. To deal with financial big data (usually unstructured), we follow the \textsf{DataOps} paradigm and implement an automatic pipeline in Fig. \ref{fig:conventional vs neofinrl}(left): task planning, data processing, training-testing-trading, and monitoring agents' performance. Through this pipeline, we continuously produce DRL benchmarks on dynamic market datasets.

In this paper, we present an openly accessible FinRL-Meta library that has been actively maintained by the AI4Finance community. We aim to create an infrastructure to enable near real-time paper trading and facilitate the real-world adoption of financial reinforcement learning. This is relevant to the broader RL research community since it provides a rare case of a task that can be tested against real-world performance without major investment, while robotics requires simulation or expensive equipment and games are available in simulations.

Fig. \ref{fig:conventional vs neofinrl}(right) shows an overview of data-driven financial reinforcement learning.  First, following a DataOps paradigm \cite{DataOps, atwal2019practical, ereth2018dataops}, we provide hundreds of market environments through an automatic pipeline that collects dynamic datasets from real-world markets and processes them into standard gym-style market environments.
Second, we reproduce popular papers as benchmarks, including high-frequency stock trading, cryptocurrency trading and stock portfolio allocation, serving as stepping stones for users to design new strategies. With the help of the data engineering pipeline, we hold our benchmarks
on cloud platforms so that users can visualize their own results and assess the relative performance via community-wise competitions. Third, FinRL-Meta provides tens of Jupyter/Python demos as educational materials, organized in a curriculum, and a documentation website to serve the rapidly growing community. 

The remainder of this paper is organized as follows. Section 2 reviews existing works. Section 3 describes challenges and presents an overview of our FinRL-Meta framework. Section 4 describes how we process data into market environments. In Section 5, we benchmark popular DRL papers. Finally, we conclude this paper in Section 6.

%% file: NeurIPS_2020_version/Section2_ExistingWorks.tex
\section{Related Works}

We review DataOps practices and existing works on data-driven reinforcement learning.

\textbf{DataOps practices}: DataOps \cite{ereth2018dataops, atwal2019practical, DataOps} applies the ideas of lean development and DevOps to the data analytics field. DataOps practices have been developed in companies and organizations to improve the quality and efficiency of data analytics \cite{atwal2019practical}. These implementations consolidate various data sources, unify and automate the pipeline of data analytics, including data accessing, cleaning, analysis, and visualization.

However, the DataOps methodology has not been applied to financial reinforcement learning researches. Most researchers access data, clean data, and extract technical indicators (features) in a case-by-case manner, which involves heavy manual work and may not guarantee the data quality.

\textbf{Data-driven reinforcement learning}: Environments are crucial for training DRL agents \cite{sutton2018reinforcement}. 
\begin{itemize}[leftmargin=*]
    \item \textbf{OpenAI gym} \cite{brockman2016openai} provides standardized environments for a collection of benchmark problems that expose a common interface, which is widely supported by many libraries \cite{stable-baselines, liang2018rllib, elegantrl}. Three trading environments, \textsf{TradingEnv, ForexEnv, and StocksEnv}, are included to support Stock and FOREX markets. However, it has not been updated for years.
    \item \textbf{D4RL} \cite{fu2020d4rl} introduces the idea of \textit{Datasets for deep data-driven reinforcement learning} (D4RL). It provides benchmarks in offline RL. However, D4RL does not provide financial environments.
    \item \textbf{FinRL} \cite{liu2020finrl,liu2021finrl} is an open-source library that builds a full pipeline for financial reinforcement learning. It contains three market environments, i.e., stock trading, portfolio allocation, crypto trading, and two data sources, i.e., Yahoo Finance and WRDS. However, those market environments of FinRL cannot meet the community's growing demands.
    \item \textbf{NeoRL} \cite{qin2021neorl} collected offline RL environments for four areas, CityLearn \cite{vazquez2019cityLearn}, FinRL \cite{liu2020finrl,liu2021finrl}, Industrial Benchmark \cite{hein2017benchmark}, and MuJoCo \cite{todorov2012mujoco}, where each area contains several gym-style environments. Regarding financial aspects, it directly borrows market environments from FinRL.
\end{itemize}

\textbf{Benchmarks of financial reinforcement learning}:
Many researches applied DRL algorithms in quantitative finance \cite{xiong2018practical, yang2020deep, zhang2020deep, ardon2021towards, amrouni2021abides, coletta2021towards} by building their own market environments. Despite the above-mentioned open-source libraries that provide some useful environments, there are no established benchmarks yet. On the other hand, the data accessing, cleaning and factor extraction processes are usually limited to data sources like Yahoo Finance and Wharton Research Data Services (WRDS).

%% file: NeurIPS_2020_version/Section3_Architecture.tex
\section{Financial Reinforcement Learning and FinRL-Meta Framework}

\definecolor{Gray}{RGB}{217,234,211}
\begin{table}
\caption{List of state space, action space, and reward function.}
\small
\renewcommand{\arraystretch}{1.2}
\centering
\begin{tabular}{|l|l|l|}
    \hline   
    \textbf{Key components} & \textbf{Attributes} 
    \\ \hline
    State & Balance ${b}_{t}\in \mathbb{R}_+$;~Shares $\bm{h}_{t}\in \mathbb{Z}_+^{n}$ \\
    & Opening/high/low/close price $\bm{o}_{t}, \bm{h}_{t}, \bm{l}_{t},\bm{p}_{t} \in \mathbb{R}_+^{n}$ \\ 
    & Trading volume $\bm{v}_{t}\in \mathbb{R}_+^{n}$ \\
    & Fundamental indicators; Technical indicators \\
    & Social data; Sentiment data \\
    &  Smart beta indexes, etc. \\ 
    \hline
     Action & Buy/Sell/Hold   \\
    & Short/Long \\
    & Portfolio weights 
    \\ \hline
     Reward &  Change of portfolio value \\  
    & Portfolio log-return \\
    & Sharpe ratio 
    \\ \hline
     Environments &  Dow-$30$, S\&P-$500$, NASDAQ-$100$ \\  
    &  Cryptocurrencies \\
    & Foreign currency and exchange \\ 
    &  Futures;~Options;~ETFs;~Forex \\ 
    & CN securities;~US securities;~NMS US securities\\
    & Paper trading; Living Trading 
    \\ \hline
\end{tabular}\vspace{0.050in}
\label{event:eventTypes}
\end{table}

We describe financial reinforcement learning and its challenges, then provide an overview of our FinRL-Meta framework.

\subsection{Financial Reinforcement Learning and Challenges}

Assuming full observability, we model a trading task as a Markov Decision Process (MDP) with five tuples \cite{sutton2018reinforcement} $(\mathcal{S}, \mathcal{A}, \mathbb{P}, r, \gamma)$, where $\mathcal{S}$ and $\mathcal{A}$ denote the state space and action space, respectively, $\mathbb{P}(s'|s,a)$ is the transition probability of an unknown environment, $r(s,a, s')$ is a reward function, and $\gamma \in (0, 1] $ is a discount factor. A trading agent learns a policy $\pi(s_t|a_t)$ that maximizes the discounted cumulative return $R = \sum^{T}_{t=0} \gamma ^t r(s_t, a_t, s_{t+1})$ over a trading period $t=0,1,...,T$.

The historical dataset before time $0$ is used to train the trading agent. Note that we process the dataset into a market environment, following the \textit{de facto} standard of OpenAI gym \cite{brockman2016openai}. In Table \ref{event:eventTypes}, we list the state space, action space, and reward function.
\begin{itemize} [leftmargin=*]
    \item \textbf{State} $s \in \mathcal{S}$: A state represents an agent's perception of a market environment, which may include balance, shares, OHLCV values, technical indicators, social data, sentiment data, etc. 
    \item \textbf{Action} $a \in \mathcal{A}$: An action is taken from the allowed action set at a state. Actions may vary for different trading tasks, e.g., for stock trading, the actions are the number of shares to buy/sell for each stock, while for portfolio allocation, the actions are the allocation weights of the capital. 
    \item \textbf{Reward} $r(s,a,s')$: Reward is an incentive mechanism for an agent to learn a better policy. Several common reward functions are provided: 1). Change of portfolio value $r(s,a,s') = v' - v$, where $v'$ and $v$ are portfolio values at state $s'$ and $s$, respectively; 2). Portfolio log return $r(s,a,s') = \log(v'/v)$; and 3). Sharpe ratio \cite{Sharpe} defined in Section \ref{sec:performance_metrics}.
\end{itemize}

The above full observability assumption can be extended to partial observation (the underlying states cannot be directly observed), i.e., partially observable Markov Decision Process (POMDP). A POMDP model utilizes a Hidden Markov Model (HMM) \cite{mamon2007hidden} to model a time series that is caused by a sequence of unobservable states. 
Considering the noisy financial data, it is natural to assume that a trading agent cannot directly observe market states. Studies suggested that the POMDP model can be solved by using recurrent neural networks, e.g., an off-policy Recurrent Deterministic Policy Gradient (RDPG) algorithm \cite{liu2020adaptive}, and a  long short-term memory (LSTM) network that encodes partial observations into a state of a reinforcement learning algorithm \cite{rundo2019deep}.

Training and testing environments based on historical data may not simulate real markets accurately due to the \textit{simulation-to-reality gap} \cite{DulacArnold2020AnEI,dulac2019challenges}, and thus a trained agent cannot be directly deployed in real-world markets. We summarize three major factors for the \textit{simulation-to-reality gap} in financial reinforcement learning as follows:
\begin{itemize} [leftmargin=*]
\item \textbf{Low signal-to-noise ratio (SNR) of financial data}: Data from different sources may contain large noise \cite{wilkman2020feasibility} such as random noise, outliers, etc. It is challenging to identify alpha signals or build smart beta indexes using noisy datasets.
\item \textbf{Survivorship bias of historical market data}: Survivorship bias is caused by a tendency to focusing on existing stocks and funds without consideration of those that are delisted \cite{brown1992survivorship}. It could lead to an overestimation of stocks and funds, which will mislead the agent.
\item \textbf{Model overfitting in backtesting stage}: Existing research mainly report backtesting results. It is possible to tune hyper-parameters and retrain the agent multiple times \footnote{There is information leakage.} to obtain better backtesting results, causing model overfitting \cite{gort2022deep,de2018advances}. 
\end{itemize}

\subsection{Overview of FinRL-Meta}

\begin{figure*}
\centering
\includegraphics[scale =0.15]{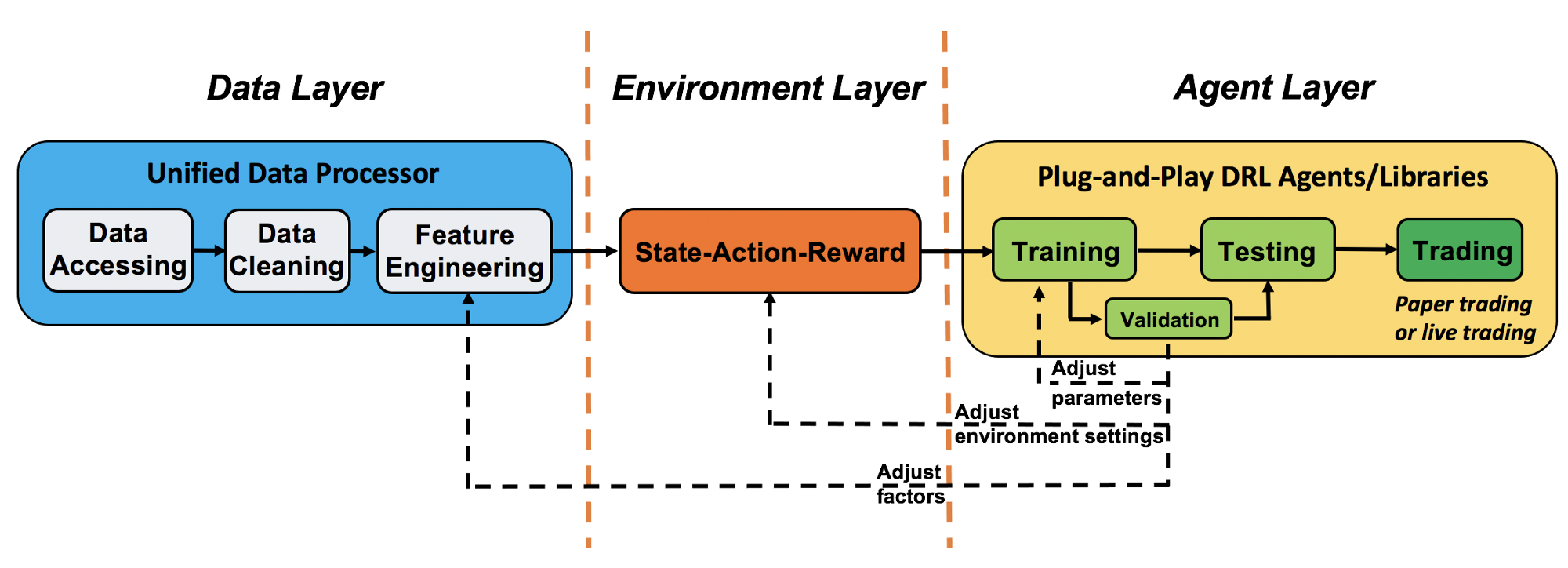}\vspace{-0.1in}
\caption{Overview of FinRL-Meta framework.}
\vspace{-0.1in}
\label{fig:finrl-meta overview}
\vspace{-2mm}
\end{figure*}

FinRL-Meta builds a universe of market environments for data-driven financial reinforcement learning. FinRL-Meta follows the \textit{de facto} standard of OpenAI Gym \cite{brockman2016openai} and the \textit{lean principle} of software development. It has the following unique features.

\textbf{Layer structure and extensibility}:
As shown in Fig. \ref{fig:finrl-meta overview}, we adopt a layered structure that consists of three layers, data layer, environment layer, and agent layer. Layers interact through end-to-end interfaces, achieving high extensibility. For updates and substitutes inside a layer, this structure minimizes the impact on the whole system.  Moreover, the layer structure allows easy extension of user-defined functions and fast updating of algorithms with high performance.

\textbf{Training-testing-trading pipeline}:
We employ a training-testing-trading pipeline that the DRL approach follows a standard end-to-end pipeline. The DRL agent is first trained in a training environment and then fined-tuned (adjusting hyperparameters) in a validation environment. Then the validated agent is tested on historical datasets (backtesting). Finally, the tested agent will be deployed in paper trading or live trading markets.

\textbf{Plug-and-play mode}: In the above training-testing-trading pipeline, a DRL agent can be directly plugged in, then trained and tested. The following DRL libraries are supported:
\begin{itemize} [leftmargin=*]
    \item \textbf{ElegantRL \cite{elegantrl}}: Lightweight, efficient and stable algorithms using PyTorch.
    \item \textbf{Stable-Baselines3 \cite{stable-baselines}}: Improved DRL algorithms based on OpenAI Baselines.
    \item \textbf{RLlib \cite{liang2018rllib}:} An open-source DRL library that offers high scalability and unified APIs. 
\end{itemize}

%% file: NeurIPS_2020_version/Section3_Environment.tex
\section{Financial Big Data and DataOps for Dynamic Datasets} \label{datasets}

\begin{figure}
\centering
\includegraphics[scale=0.45]{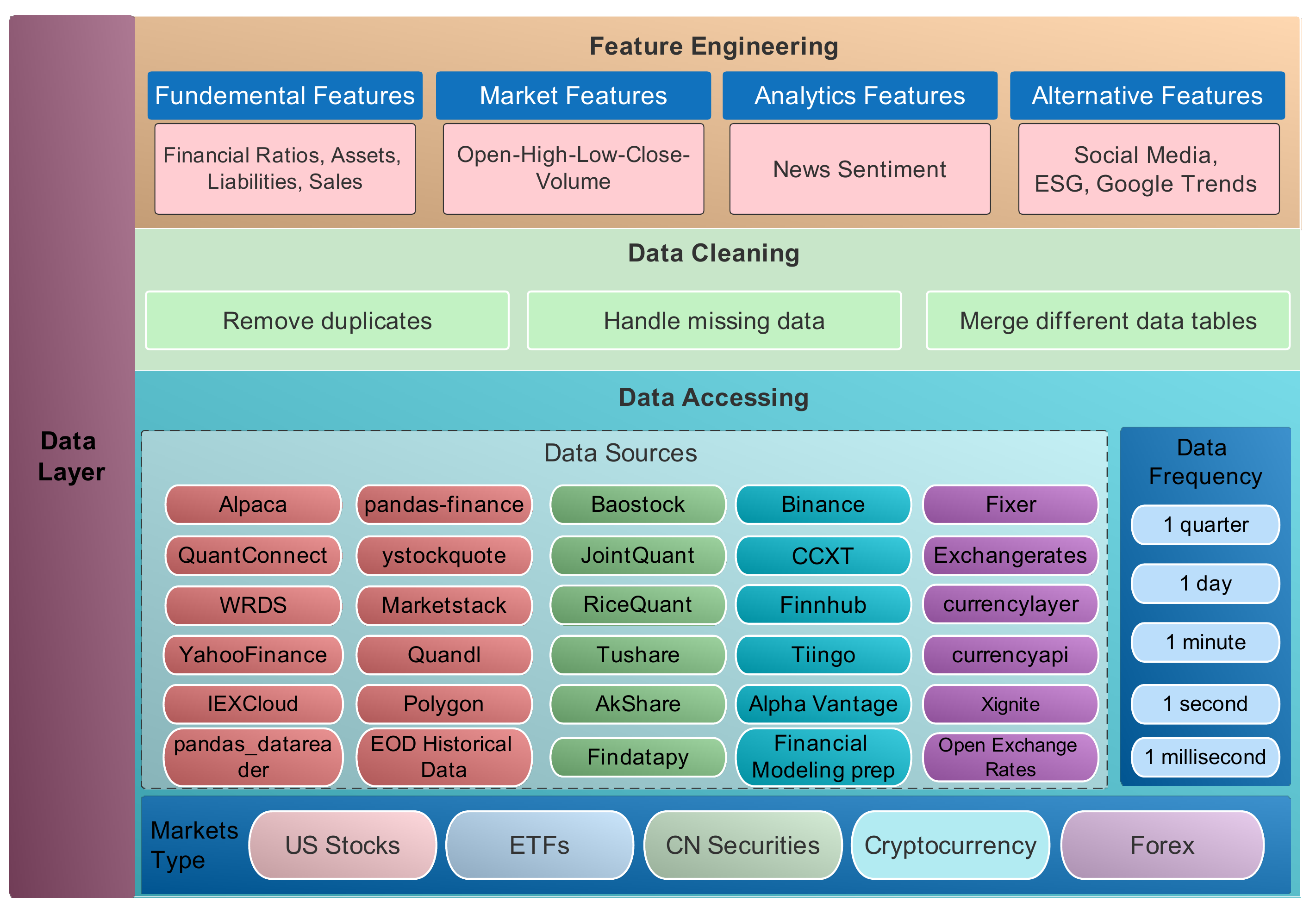}
\caption{Data layer of FinRL-Meta.}
\label{fig:datalayer}
\vspace{-2mm}
\end{figure}

Financial big data is usually unstructured in shape and form. We process four types of data \cite{de2018advances} into market environments, including fundamental data (e.g., earning reports), market data (e.g., OHLCV data), analytics (e.g., news sentiment), and alternative data (e.g., social media data, ESG data).

\subsection{Data Layer for Unstructured Financial Big Data}

\textbf{Automated pipeline for data-driven financial reinforcement learning}:  We follow the DataOps paradigm \cite{ereth2018dataops, atwal2019practical, DataOps} in the data layer. As shown in Fig. \ref{fig:datalayer}, we establish a standard pipeline for financial data engineering, which processes data from different sources into a unified market environment following the \textit{de facto} standard of OpenAI gym \cite{brockman2016openai}. We automate this pipeline with a data processor that implements the following functionalities:
\begin{itemize} [leftmargin=*]
\item \textbf{Data accessing}:
Users can connect data APIs of different market platforms via our common interfaces. Users can access data agilely by specifying the start date, end date, stock list, time interval, and other parameters. FinRL-Meta has supported more than $30$ data sources, covering stocks, cryptocurrencies, ETFs, forex, etc. 
\item \textbf{Data cleaning}:
Raw data retrieved from different data sources are usually of various formats and with erroneous or missing data to different extents. It makes data cleaning highly time-consuming. With a data processor, we automate the data cleaning process. In addition, we use stock ticker name and data frequency as unique identifiers to merge all types of data into one unified data table.
\item \textbf{Feature engineering}: In feature engineering, FinRL-Meta aggregates effective features which can help improve model predictive performance. We provide various types of features, including but not limited to fundamental, market, analytics, and alternative features. Users can quickly add features using open-source libraries or add user-defined features. Users can add new features in two ways: 1) Write a user-defined feature extraction function directly. The returned features are added to a feature array. 2) Store the features in a file, and put it in a default folder. Then, an agent can read these features from the file.
\end{itemize}

\textbf{Automated feature engineering}: 
FinRL-Meta currently supports four types of features:
\begin{itemize} [leftmargin=*]
\item \textbf{Fundamental features}:
Fundamental features are processed based on the earnings data in SEC filings queried from WRDS. The data frequency is low, typical quarterly, e.g., four data points in a year. To avoid information leakage, we use a two-month lag beyond the standard quarter end date, e.g., Apple released its earnings report on 2022/07/28 for the third quarter (2022/06/25) of year 2022. Thus for the quarter between 04/01 and 06/30, our trade date is adjusted to 09/01 (same method for other three quarters). We also provide functions in our data processor for calculating financial ratios based on earnings data such as earnings per share (EPS), return on asset (ROA), price to earnings (P/E) ratio, net profit margin, quick ratio, etc.

\item \textbf{Market features}:
 Open-high-low-close price and volume data are the typical market data we can directly get from querying the data API. They have various data frequencies, such as daily prices from YahooFinance, TAQ (Millisecond Trade and Quote) from WRDS. In addition, we automate the calculation of technical indicators based on OHLCV data by connecting the Stockstats\footnote{Github repo: \url{https://github.com/jealous/stockstats}} or TA-lib library\footnote{Github repo: \url{https://github.com/mrjbq7/ta-lib}} in our data processor, such as Moving Average Convergence Divergence (MACD), Average Directional Index (ADX), Commodity Channel Index (CCI), etc.

\item \textbf{Analytics features}:
We provide news sentiment for analytics features. First, we get the news headline and content from WRDS \cite{xinyi_2019}. Next, we use NLTK.Vader\footnote{Github repo: \url{https://github.com/nltk/nltk}} to calculate sentiment based on the sentiment compound score of a span of text by normalizing the emotion intensity (positive, negative, neutral) of each word. For the time alignment with market data, we use the exact enter time, i.e., when the news enters the database and becomes available, to match the trade time. For example, if the trade time is every ten minutes, we collect the previous ten minutes' news based on the enter time; if no news is detected, then we fill the sentiment with 0.
\item \textbf{Alternative features}:
Alternative features are useful, but hard-to-obtain from different data sources \cite{de2018advances}, such as ESG data, social media data, Google trend searches, etc. ESG (Environmental, social, governance) data are widely used to measure the sustainability and societal impacts of an investment. The ESG data we provide is from the Microsoft Academic Graph database, which is an open resource database with records of scholar publications. We have functions in our data processor to extract AI publication and patent data, such as paper citations, publication counts, patent counts, etc. We believe these features reflect companies' research and development capacity for AI technologies \cite{fang2019practical,chen2020quantifying}. It is a good reflection of ESG research commitment.
\end{itemize}

\subsection{Environment Layer for Creating Dynamic Market Environments}
\label{sect:env_layer}

FinRL-Meta follows the OpenAI gym-style \cite{brockman2016openai} to create market environments using the cleaned data from the data layer. It provides hundreds of environments with a common interface. Users can build their environments using FinRL-Meta's interfaces, share their results and compare a strategy's trading performance. Following the gym-style \cite{brockman2016openai}, each environment has three functions as follows:
\begin{itemize}[leftmargin=*]
    \item \texttt{reset()} function resets the environment back to the initial state $s_0$
    \item \texttt{step()} function takes an action $a_t$ from the agent and updates state from $s_t$ to $s_{t+1}$.
    \item \texttt{reward()} function computes the reward value transforming from $s_t$ to $s_{t+1}$ by action $a_t$.
\end{itemize}
Detailed descriptions can be found in \cite{yang2020deep}\cite{gort2022deep}.

We plan to add more environments for users' convenience. For example, we are actively  building market simulators using Limit-order-book data \footnote{Github repo: \url{https://github.com/AI4Finance-Foundation/Market_Simulator}}, where we simulate the market from the playback of historical limit-order-book-level data and an order matching mechanism. We foresee the flexibility and potential of using a Hidden Markov Model (HMM) \cite{mamon2007hidden}  or a generative adversarial net (GAN) \cite{goodfellow2014generative} to generate market scenarios \cite{coletta2021towards}.

\textbf{Incorporating trading constraints to model market frictions}:
To better simulate real-world markets, we incorporate common market frictions (e.g., transaction costs and investor risk aversion) and portfolio restrictions (e.g., non-negative balance). 
\begin{itemize}[leftmargin=*]
\item \textbf{Flexible account settings}: Users can choose whether to allow buying on margin or short-selling.
\item \textbf{Transaction cost}: We incorporate the transaction cost to reflect market friction, e.g., $0.1\%$ of each buy or sell trade.
\item \textbf{Risk-control for market crash}: In FinRL \cite{liu2020finrl,liu2021finrl}, a turbulence index \cite{kritzman2010skulls} is used to control risk during market crash situations. However, calculating the turbulence index is time-consuming. It may take minutes, which is not suitable for paper trading and live trading. We replace the financial turbulence index  with the volatility index (VIX) \cite{whaley2009understanding} that can be accessed immediately.
\end{itemize}

\textbf{Multiprocessing training via vector environment}:
We utilize GPUs for multiprocessing training, namely, the vector environment technique of Isaac Gym \cite{makoviychuk2021isaac}, which significantly accelerates the training process.  In each CUDA core, a trading agent interacts with a market environment to produce transitions in the form of $\{$state, action, reward, next state$\}$. Then, all the transitions are stored in a replay buffer and later used to update a learner. By adopting this technique, we successfully achieve the multiprocessing simulation of hundreds of market environments to improve the performance of DRL trading agents on large datasets.

\subsection{Advantages}
Our DataOps pipeline is automatic, which gives us the following three advantages.

\textbf{Curriculum for newcomers}: We provide an educational curriculum, as shown in Fig. \ref{fig:tutorials},
for community newcomers with different levels of proficiency and learning goals. Users can grow programming skills by gradually changing the data/environment layer following instructions on our website.

\textbf{Benchmarks on cloud}: We provide demos on a cloud platform, Weights \& Biases \footnote{Website: \url{https://wandb.ai/site}}, to demonstrate the training process. We define the hyperparameter sweep, training function, and initialize an agent to train and tune hyperparameters. On the cloud platform Weights \& Biases, users are able to visualize their results and assess  the relative performance via community-wise competitions.

\textbf{Curriculum learning for agents}: Based on FinRL-Meta (a universe of market environments, say $\geq 100$), one is able to construct an environment by sampling data samples from multiple market datasets, similar to XLand \cite{team2021open}. In this way, one can apply the curriculum learning method \cite{team2021open} to train a generally capable agent for several financial tasks.

%% file: NeurIPS_2020_version/Section4_Evaluation.tex
\section{Tutorials and Benchmarks of Financial Reinforcement Learning} \label{benchmarks}

\begin{figure}
\centering
\includegraphics[scale=0.2]{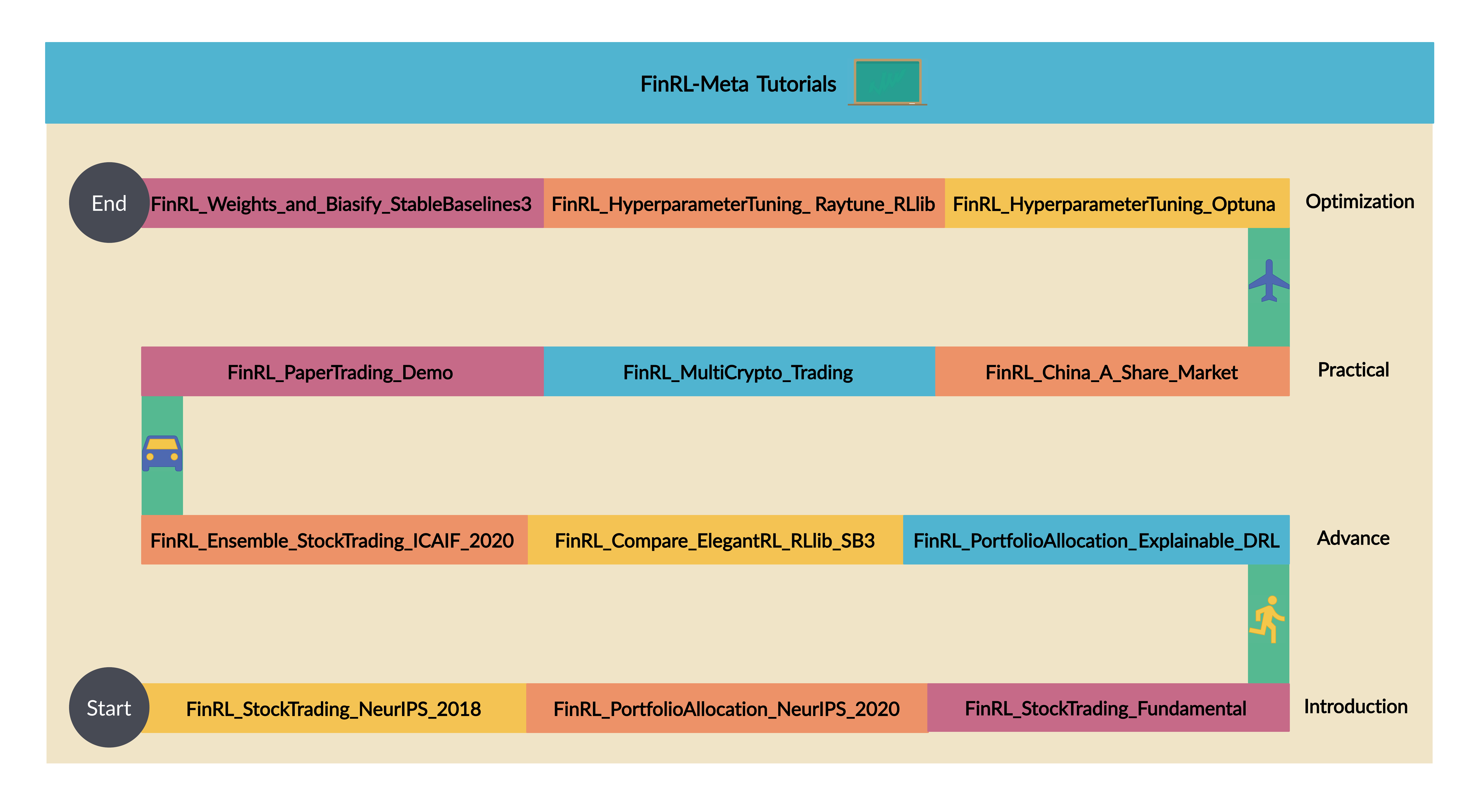}
\caption{Demos of FinRL-Meta, organized in a curriculum structure.}
\label{fig:tutorials}
\vspace{-2mm}
\end{figure}

We provide tens of tutorial notebooks to serve as stepping stones for newcomers and reproduce popular papers as benchmarks for follow-up research.

\subsection{Metrics and Baselines for Evaluating Performance}
\label{sec:performance_metrics}

We provide the following metrics to measure the trading performance: 
\begin{itemize}[leftmargin=*]
  \item \textbf{Cumulative return} $R = \frac{v - v_0}{v_0}$, where $v$ is the final portfolio value, and $v_0$ is the original capital.
  \item \textbf{Annualized return} $r = (1+R)^\frac{365}{t}-1$, where $t$ is the number of trading days.
  \item \textbf{Annualized volatility} ${\sigma}_a = \sqrt{\frac{\sum_{i=1}^{n}{(r_i-\bar{r})^2}}{n-1}}$, where $r_i$ is the annualized return in year $i$, $\bar{r}$ is the average annualized return, and $n$ is the number of years.
  \item \textbf{Sharpe ratio} \cite{Sharpe} 
$S_T = \frac{\text{mean}(R_t) - r_f}{\text{std}(R_t)}$, where $R_t = \frac{v_t - v_{t-1}}{v_{t-1}}$, $r_f$ is the risk-free rate, and $t=1,...,T$.
  \item \textbf{Max. drawdown}: The maximal percentage loss in portfolio value.
\end{itemize}

The following baseline trading strategies are provided for comparisons:
\begin{itemize}[leftmargin=*]
    \item \textbf{Passive trading strategy} \cite{malkiel2003passive} is a well-known long-term strategy. The investors just buy and hold selected stocks or indexes without further activities.
    \item \textbf{Mean-variance and min-variance strategy} \cite{ang2012mean} are two widely used strategies that look for a balance between risks and profits. They select a diversified portfolio in order to achieve higher profits at a lower risk.
    \item \textbf{Equally weighted strategy} is a portfolio allocation strategy that gives equal weights to different assets, avoiding allocating overly high weights on particular stocks. 
\end{itemize}

\subsection{Tutorials and Demos in Jupyter Notebooks}

For educational purposes, we provide Jupyter notebooks as tutorials\footnote{https://github.com/AI4Finance-Foundation/FinRL-Tutorials} to help newcomers get familiar with the whole pipeline.
\begin{itemize}[leftmargin=*]
    \item \textbf{Stock trading} \cite{xiong2018practical}: We apply popular DRL algorithms to trade multiple stocks.
    \item \textbf{Portfolio allocation} \cite{liu2020finrl}: We use DRL agents to optimize asset allocation in a set of stocks.
    \item \textbf{Cryptocurrency trading} \cite{liu2020finrl}: We reproduce the experiment \cite{liu2020finrl} on $10$ popular cryptocurrencies.
    \item \textbf{Multi-agent RL for liquidation strategy analysis} \cite{bao2019multiagent}: We reproduce the experiment in \cite{bao2019multiagent}. The multi-agent optimizes the shortfalls in the liquidation task, which is to sell given shares of one stock sequentially within a given period, considering the costs arising from the market impact and the risk aversion.
    \item \textbf{Ensemble strategy for stock trading} \cite{yang2020deep}: We reproduce the experiment in \cite{yang2020deep} that employed an ensemble strategy of several DRL algorithms on the stock trading task.
    \item \textbf{Paper trading demo}: We provide a demo for paper trading. Users could combine their own strategies or trained agents in paper trading.
    \item \textbf{China A-share demo}: We provide a demo based on the China A-share market data.
    \item \textbf{Hyperparameter tuning}: We provide several demos for hyperparameter tuning using Optuna \cite{akiba2019optuna} or Ray Tune \cite{liaw2018tune}, since hyperparameter tuning is critical for better performance.
\end{itemize}

\begin{figure}
\centering
\includegraphics[scale=0.51]{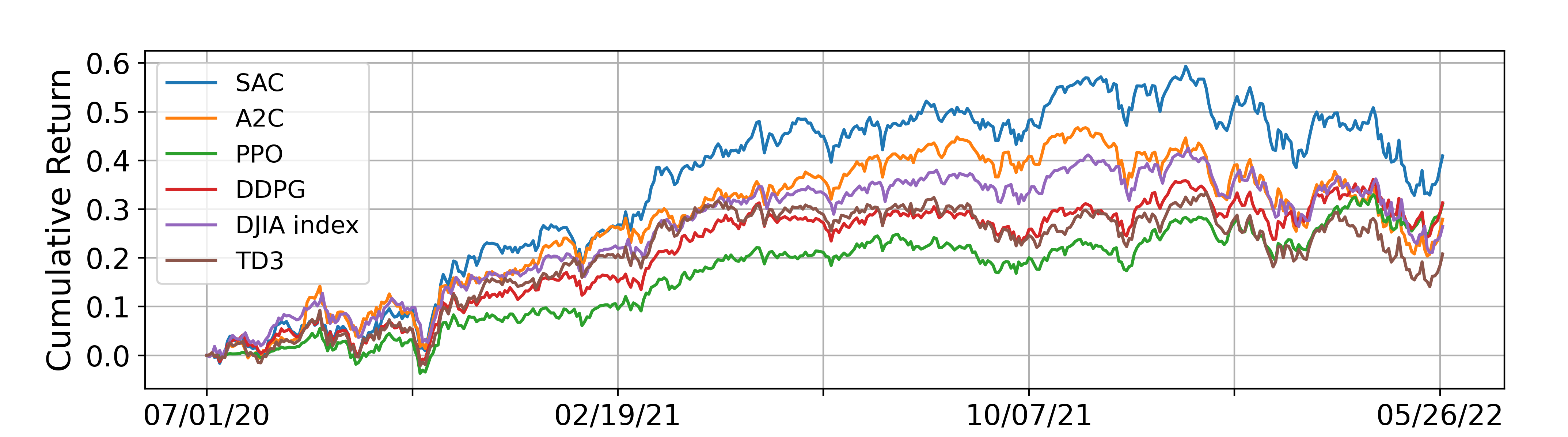}
\caption{Reproducing stock trading (left) of \cite{xiong2018practical}.}
\label{fig:stock_trading_performance}
\vspace{-1mm}
\end{figure}

\subsection{Reproducing Prior Papers as Benchmarks}

We have reproduced experiments in several papers as benchmarks. Users can study our codes for research purpose or use them as stepping stones for deploying trading strategies in live markets. In this subsection, we introduce three home-grown examples specifically. For more benchmarks, please refer to Appendix.  \ref{sec:appendixB}.

\textbf{Stock trading task} \cite{xiong2018practical}: We access Yahoo! Finance database and select the 30 constituent stocks (accessed at 07/01/2020) in Dow Jones Industrial Average (DJIA). We use data from 01/01/2009 to 06/30/2020 for training and data from 07/01/2020 to 05/31/2022 for testing. We use technical indicators in our state space, e.g., Moving Average Convergence Divergence (MACD),  Relative Strength Index (RSI), Commodity Channel Index (CCI), Average Directional Index (ADX), etc.

As shown in Fig.~\ref{fig:stock_trading_performance} (left), we train five popular DRL algorithms to trade and compare their results with the DJIA index. We show a detailed walkthrough of how DRL works in the stock trading task, on which many subsequent works are based \cite{xiong2018practical}. This benchmark is beneficial for getting into the field of RL in finance. 

\textbf{Podracer on the cloud \cite{finrl_podracer_2021,liu2021podracer}}: We reproduce cloud solutions of population-based training, e.g., generational evolution \citep{finrl_podracer_2021} and tournament-based evolution \citep{liu2021podracer}. If GPUs are abundant, users can take advantage of this benchmark to meet the real-time requirement of high-frequency trading tasks. Detailed instructions are provided on our website.

\textbf{Ensemble strategy} \cite{yang2020deep}: The ensemble method combines different agents to obtain an adaptive one, which inherits the best features of the agents and performs remarkably well in practice. We consider three component algorithms, Proximal Policy Optimization (PPO), Advantage Actor-Critic (A2C), and Deep Deterministic Policy Gradient (DDPG), which have different strengths and weaknesses. For instance, A2C is good at dealing with a bearish trend market. PPO is good at following trends and acts well in generating more returns in a bullish market. DDPG can be used as a complementary strategy to PPO in a bullish trend. Using a rolling window, an ensemble agent automatically selects the best model for each test period. Again on the 30 constituent stocks of the DJIA index, we use data from 04/01/2009 to 06/30/2019 for training, and data from 07/01/2020 to 03/31/2022 for validation and testing through a quarterly rolling window.

\begin{figure}
\centering
\includegraphics[scale=0.51]{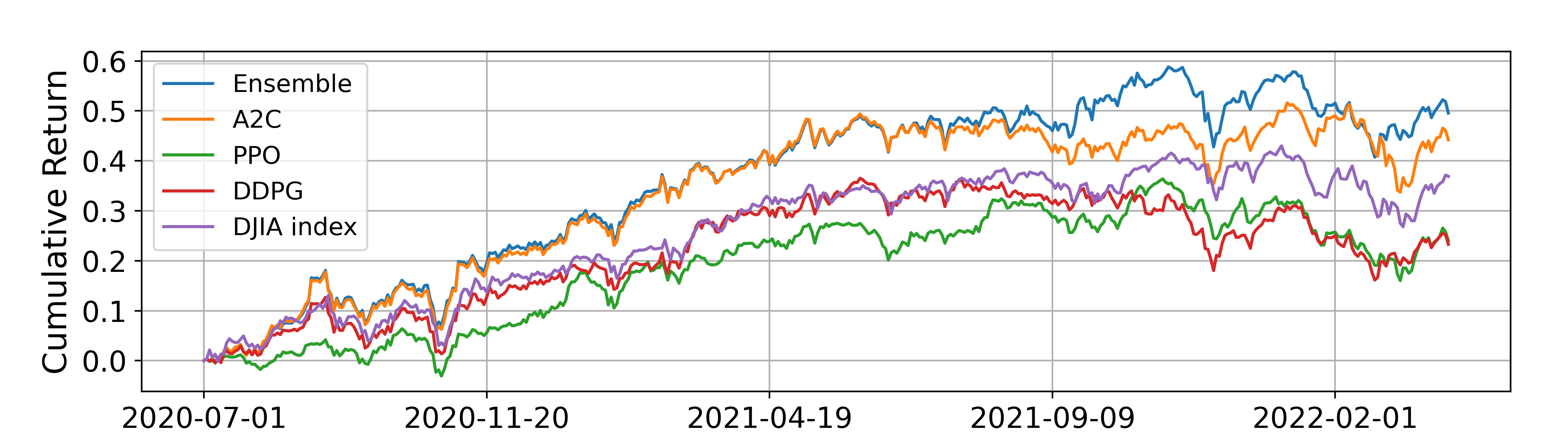}
\caption{Reproducing the ensemble strategy of \cite{yang2020deep}: cumulative return.}
\label{fig:compare_returns_ensemble}
\vspace{-1mm}
\end{figure}

From Fig.~\ref{fig:compare_returns_ensemble} and Table~\ref{tab:ensemble_performance}, we observe that the ensemble agent outperforms other agents. In the experiment, the ensemble agent has the highest Sharpe ratio of $1.53$, which means it performs the best in balancing risks and profits. This benchmark demonstrates that the ensemble strategy is effective in constructing a more reliable agent based on several component DRL agents.\looseness=-1

\begin{table*}[t]
 \centering
 \renewcommand{\arraystretch}{1.3}
\resizebox{1\textwidth}{!}{
\begin{tabular}{|c| c |c |c |c |c |}
\hline
(2020/07/01-2022/03/31) & Ensemble \cite{yang2020deep} & A2C & PPO & DDPG & DJIA index \\
\hline
Annual Return & 25.9\% & 23.3\% & 13.1\% & 12.7\% & 19.7\% \\
\hline
Annual Volatility & 15.9\% & 16.2\% & 13.4\% & 15.0\% & 14.4\% \\
\hline
Sharpe Ratio & 1.53 & 1.37 & 0.99 & 0.88 & 1.32 \\
\hline
Calmar Ratio & 2.27 & 1.97 & 0.88 & 0.85 & 1.74 \\
\hline
Max Drawdown & -11.4\% & -11.8\% & -14.9\% & -14.9\% & -11.3\% \\
\hline
\end{tabular}}
\caption{Reproducing the ensemble strategy of \cite{yang2020deep}.}
\label{tab:ensemble_performance}
\vspace{-2mm}
\end{table*}

%% file: NeurIPS_2020_version/appendix.tex

\section*{Checklist}

\begin{enumerate}

\item For all authors...
\begin{enumerate}
  \item Do the main claims made in the abstract and introduction accurately reflect the paper's contributions and scope?
  \item Did you describe the limitations of your work? More computational cost.
  \item Did you discuss any potential negative societal impacts of your work? May lead to future works with higher computational cost.
  \item Have you read the ethics review guidelines and ensured that your paper conforms to them?
\end{enumerate}

\item If you are including theoretical results...
\begin{enumerate}
  \item Did you state the full set of assumptions of all theoretical results?
  \item Did you include complete proofs of all theoretical results?
\end{enumerate}

\item If you ran experiments...
\begin{enumerate}
  \item Did you include the code, data, and instructions needed to reproduce the main experimental results (either in the supplemental material or as a URL)?
  \item Did you specify all the training details (e.g., data splits, hyperparameters, how they were chosen)?
        \item Did you report error bars (e.g., with respect to the random seed after running experiments multiple times)?
        \item Did you include the total amount of compute and the type of resources used (e.g., type of GPUs, internal cluster, or cloud provider)?
\end{enumerate}

\item If you are using existing assets (e.g., code, data, models) or curating/releasing new assets...
\begin{enumerate}
  \item If your work uses existing assets, did you cite the creators?
  \item Did you mention the license of the assets?
  \item Did you include any new assets either in the supplemental material or as a URL?
  \item Did you discuss whether and how consent was obtained from people whose data you're using/curating?
  \item Did you discuss whether the data you are using/curating contains personally identifiable information or offensive content?
\end{enumerate}

\item If you used crowdsourcing or conducted research with human subjects...
\begin{enumerate}
  \item Did you include the full text of instructions given to participants and screenshots, if applicable?
  \item Did you describe any potential participant risks, with links to Institutional Review Board (IRB) approvals, if applicable?
  \item Did you include the estimated hourly wage paid to participants and the total amount spent on participant compensation?
\end{enumerate}

\end{enumerate}


\newpage
\appendix

\definecolor{Gray}{RGB}{217,234,211}
\begin{table}
\caption{List of key terms for reinforcement learning.}
\small
\renewcommand{\arraystretch}{1.2}
\centering
\begin{tabular}{|l|l|l|}
   \hline 
    \textbf{Key Terms} & \textbf{Description} \\
    \hline
    Agent \cite{sutton2022quest} 
    & A decision maker\\
    \hline
    Environment \cite{sutton2022quest}
    & A world with which an agent interacts with\\
    \hline
    Gym-style environment \cite{brockman2016openai}
    & A standard form of DRL environment by OpenAI \\
    \hline
    Markov Decision Process (MDP)
    & A mathematical framework to model decision-making problems\\
    \hline
    State, Action, Reward 
    & Three main factors in an agent-environment interaction \\
    \hline
    Policy
    & A rule that agent follow to make decision\\
    \hline
    Policy gradient & An approach to solve RL problems by optimizing the policy directly\\
    \hline
    Deep Q-Learning (DQN) \cite{DQN} &  The first DRL algorithm that uses a neural network to approximate the Q-function \\
    \hline
    DDPG \cite{DDPG} & Deep Deterministic Policy Gradient \\
    \hline
    PPO \cite{PPO_2017} & Proximal Policy Optimization \\
    \hline
    Hyperparameter tuning
    & Change hyperparameter during training to get a converged result faster \\
    \hline
    Ensemble strategy
    & An ML technique. Here we combine several DRL agent to a better model\\
    \hline
    Population-based training (PBT) & Optimise a population of models and hyperparameters, and select the optimal set\\
    \hline
    Generational evolution \cite{finrl_podracer_2021} & Employing an evolution strategy over generations \\
    \hline 
    Tournament-based evolution \cite{liu2021podracer} & An evolution by asynchronously updating a tournament board of models \\
    \hline
    Curriculum learning &  Train an ML model from easier to harder data, imitating the human curriculum\\
    \hline
    Simulation-to-reality gap 
    &  The difference between simulation environment and real-world task\\
    \hline
\end{tabular}
\label{table:RL_terms}
\end{table}

\begin{table}
\caption{List of key terms for finance.}
\small
\centering
\begin{tabular}{|l|l|l|}
   \hline 
    \textbf{Key Terms} & \textbf{Description} \\
    \hline
    Algorithmic trading & A method of trading using designed algorithm instead of human traders\\
    \hline
    Backtesting
    & A method to see how a strategy performs on a certain period of historical data\\
    \hline
    Signal-to-noise ratio (SNR)
    & A ratio of desired signal (good data) to undesired signal (noise)\\
    \hline
    DataOps & A series of principles and practices to improve the quality of data science\\
    \hline
    Sentiment data 
    & A category in financial big data that contains subjective viewpoints\\
    \hline
    Historical data
    & All kinds of data that already existed in the past\\
    \hline
    Survivorship bias
    & A bias caused by only seeing existed examples, but not those already died out\\
    \hline
    Information leakage
    & When the data contains future information, causing model overfitting\\
    \hline
    Paper trading 
    & Simulation of buying and selling without using real money\\
    \hline
    OHLCV
    & A popular form of market data with: Open, High, Low, Close, Volume\\
    \hline
    Technical indicators & A statistical calculation based on OHLCV data to indicate future price trends \\
    \hline
    Market frictions & A financial market friction as anything that interferes with trade.\\
    \hline
    Market crash
    & A huge drop of market price within a very short time\\
    \hline
    Volatility index (VIX) \cite{whaley2009understanding}
    & A market index that shows the market's expectations for volatility\\
    \hline
    Limit Order Book (LOB) & A list to record the interest of buyers and sellers\\
    \hline
    Smart beta index & An enhanced indexing strategy to beat a benchmark index\\
    \hline
    Liquidation, trade execution & An investor closes their position in an asset\\
    \hline
\end{tabular}
\label{table:Finance_terms}
\end{table}

\normalsize

\section{Terminology of Reinforcement Learning and Finance}

We provide a list of key terms and corresponding descriptions for reinforcement learning and finance in Table~\ref{table:RL_terms} and Table~\ref{table:Finance_terms}.

For terminologies of reinforcement learning, interested users can refer to \cite{sutton2022quest} or the classic textbook \cite{sutton2018reinforcement}. Also, the webpage\footnote{OpenI SpinningUp: \url{https://spinningup.openai.com/en/latest/spinningup/rl_intro.html}} explains key concepts of RL.

For terminologies of finance, interested users can refer to \cite{de2018advances}.

\section{DataOps Paradigm for Financial Big Data}

The DataOps paradigm~\cite{ereth2018dataops}, or more accurately the methodology, is a way of organizing people, processes and technology to deliver reliable and high quality data quickly to all its users. It helps reduce the cycle time of data engineering and improves data quality. The practice of DataOps focuses on enabling collaboration across the organization to drive agility, speed of delivery and new data initiatives. By leveraging the power of automation, DataOps aims to address the challenges associated with inefficiencies in access, preparation, integration and availability of data.

However, the DataOps approach has not been applied to research in financial reinforcement learning. Most researchers acquire data, clean it and extract technical indicators (features) on a case-by-case manner, which involves heavy manual work and may not guarantee high data quality. 

To handle the unstructured financial big data, FinRL-Meta follows the DataOps paradigm and implement an automated pipeline in Fig. \ref{fig:conventional vs neofinrl} (left): task planning, data processing, training-testing-trading, and monitoring the performance of the agents. 

\begin{itemize}
    \item The first step is task planning, such as stock trading, portfolio allocation, cryptocurrency trading, etc.
    \item Then, we do data processing, including data accessing and cleaning, and feature engineering.
    \item Next step is where RL takes part in. In particular, the training-testing-trading process in Fig. \ref{fig:finrl-meta timeline}.
    \item The final step is performance monitoring.
\end{itemize}
Through this pipeline, FinRL-meta replaces this process with a single step and can continuously process dynamic market datasets into market environments.

\section{FinRL: Financial Reinforcement Learning}\label{sec:appendixB}

\begin{figure}[t]
\centering
\includegraphics[width=4.5in]{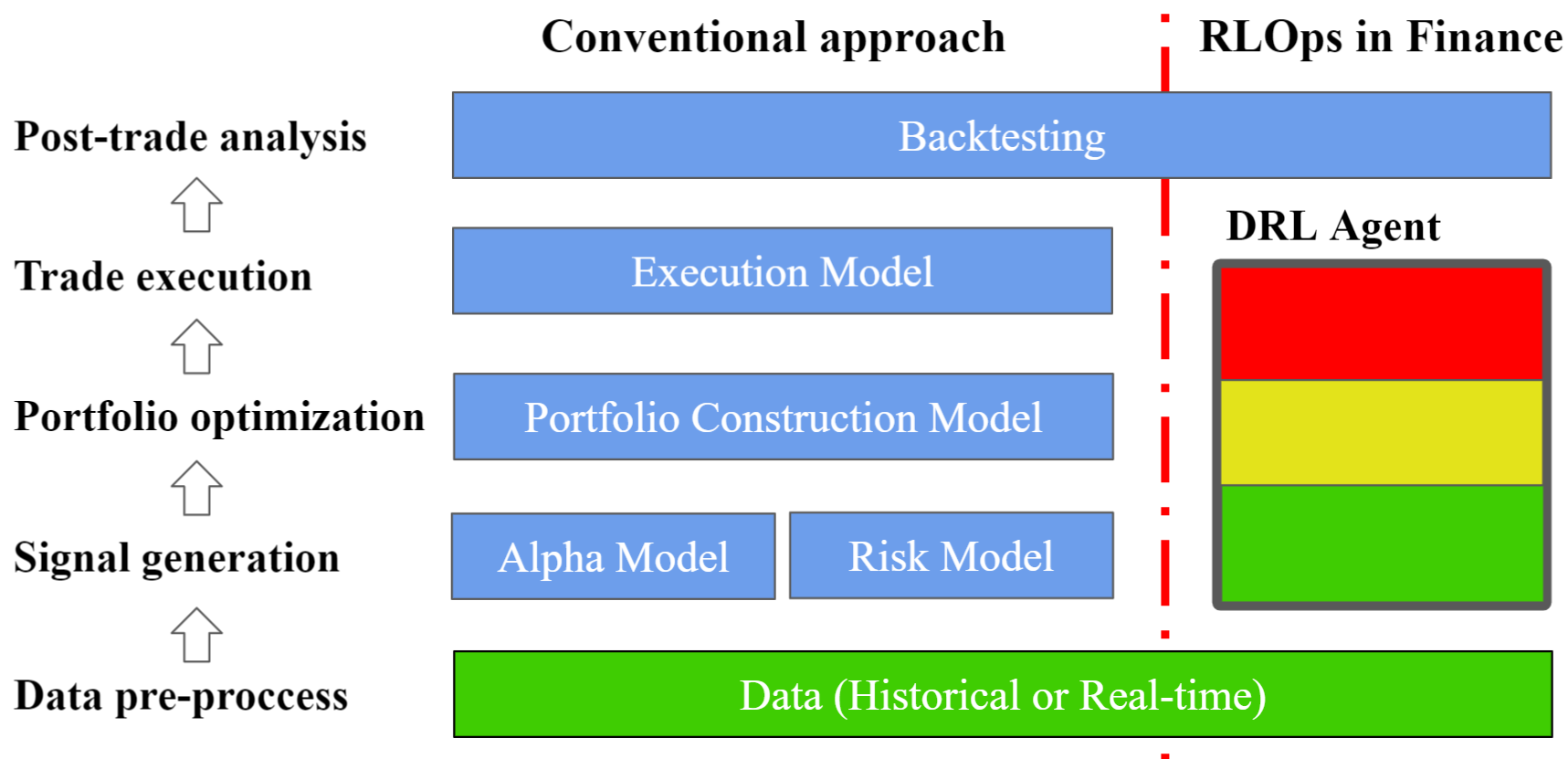}
\caption{Comparison between conventional machine learning approach and RLOps in finance for an algorithmic trading process. (This figure is from \cite{finrl_podracer_2021}.)}
\vspace{-0.2in}
\label{fig_structure}
\end{figure}

In this appendix, we take a practical perspective to provide an overview of financial reinforcement learning. First, we explain the paradigm of \textit{RLOps in fiance} \cite{finrl_podracer_2021} that may help deploy and maintain RL trading agents in real-world markets reliably and efficiently. Then, we selectively describe several applications along with practical demos. For applications of RL in finance, interested readers are suggested to read \cite{hambly2021recent} that provides a complete survey of various applications, including optimal execution, portfolio optimization, option pricing and hedging, market making, smart order routing, and robo-advising.

\subsection{RLOps in Finance Paradigm}

Algorithmic trading \cite{Treleaven2013AlgorithmicTR, Nuti2011AlgorithmicT} has been widely adopted in financial investments. The lifecycle of a conventional machine learning strategy may include five general stages, as shown in Fig. \ref{fig_structure} (left), namely data pre-processing, modeling and trading signal generation, portfolio optimization, trade execution, and post-trade analysis. Recently, deep reinforcement learning (DRL) \cite{silver2016mastering, silver2017mastering, sutton2018reinforcement} has been recognized as a powerful approach for quantitative finance, since it has the potential to overcome some important limitations of supervised learning, such as the difficulty in label specification and the gap between modeling, positioning, and order execution. 

We advocate extending the principle of \textit{MLOps} \cite{alla2021mlops}\footnote{MLOps is an ML engineering culture and practice that aims at unifying ML system development (Dev) and ML system operation (Ops).} to the \textit{RLOps in finance} paradigm that implements and automates the continuous training (CT), continuous integration (CI), and continuous delivery (CD) for trading strategies. We argue that such a paradigm has vast profits potential from a broadened horizon and fast speed, which is critical for wider DRL adoption in real-world financial tasks. The \textit{RLOps in finance} paradigm, as shown in Fig. \ref{fig_structure} (right), integrates middle stages (i.e., modeling and trading signal generation, portfolio optimization, and trade execution) into a DRL agent. Such a paradigm aims to help quantitative traders develop an end-to-end trading strategy with a high degree of automation, which removes the latency between stages and results in a compact software stack. The major benefit is that it can explore the vast potential profits behind the large-scale financial data, exceeding the capacity of human traders; thus, the trading horizon is lifted into a potentially new dimension. Also, it allows traders to continuously update trading strategies, which equips traders with an edge in a highly volatile market. However, the large-scale financial data and fast iteration of trading strategies bring imperative challenges in terms of computing power.

\subsection{Stock Trading}

Referring to \cite{xiong2018practical} and \cite{jiang_2017}, we access Yahoo! Finance database and select $30$ constituent stocks (accessed at 07/01/2020) in the Dow Jones Industrial Average (DJIA) index. We use data from 01/01/2009 to 06/30/2020 for training and data from 07/01/2020 to 05/31/2022 for testing. The following technical indicators are used in the state space: Moving Average Convergence Divergence (MACD), Relative Strength Index (RSI), Commodity Channel Index (CCI), Average Directional Index (ADX), etc.

Code available at: \url{https://github.com/AI4Finance-Foundation/FinRL/blob/master/tutorials/1-Introduction/FinRL_StockTrading_Fundamental.ipynb}

\begin{figure}
\centering
\includegraphics[scale=0.55]{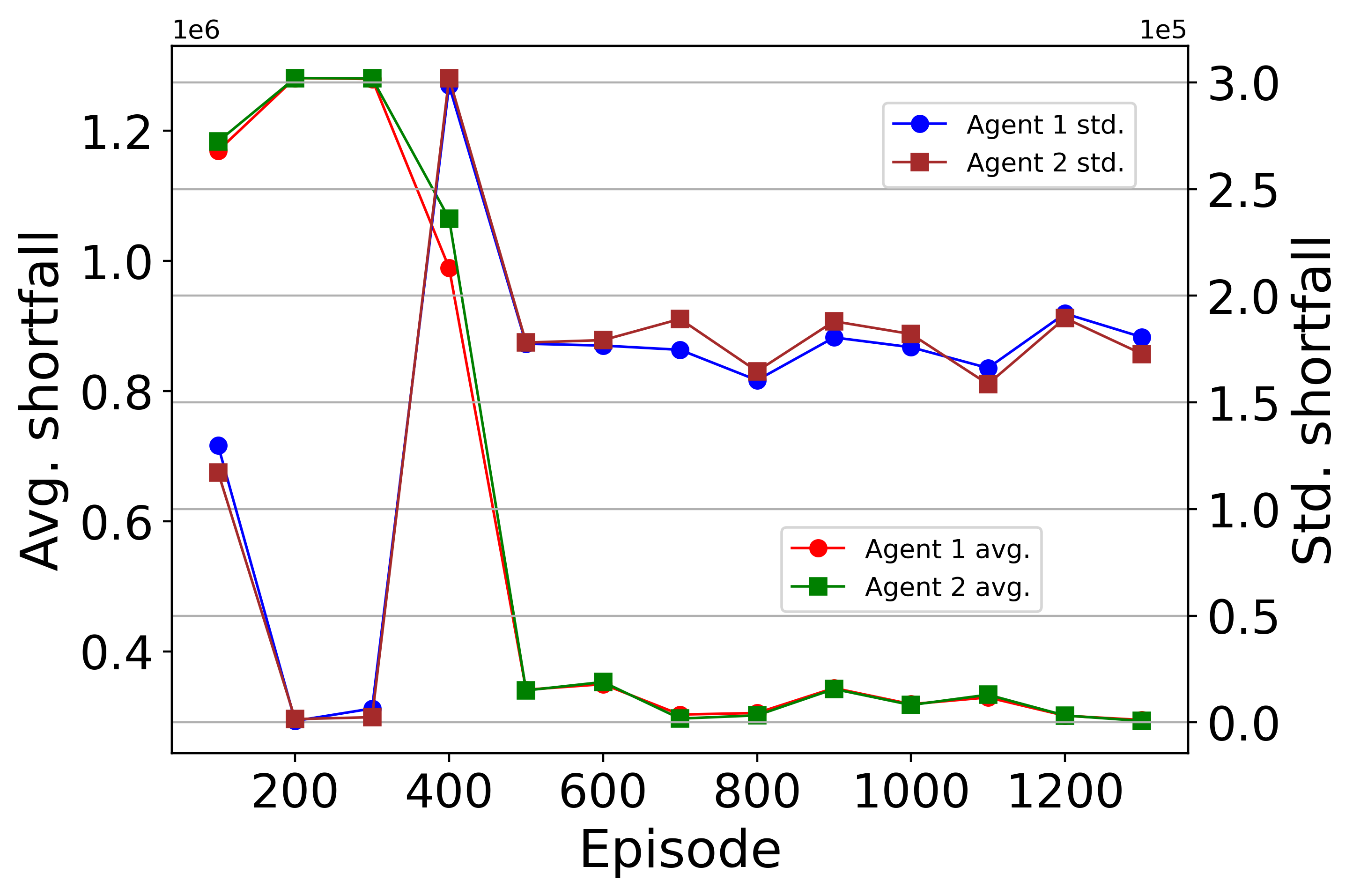}
\caption{Liquidation analysis of \cite{bao2019multiagent}.}
\label{fig:liquid_performance}
\vspace{-4mm}
\end{figure}

\subsection{Liquidation Analysis and Trade Execution}

Reproducing \cite{bao2019multiagent}, We build a simulated environment of stock prices according to the Almgren and Chriss model. Then we implement the multi-agent DRL algorithms for both competing and cooperative liquidation strategies. This benchmark demonstrates the trade execution task using deep reinforcement learning algorithms. When trading, traders want to minimize the expected trading cost, which is also called implementation shortfall. In Fig.~\ref{fig:liquid_performance}, there are two agents, and we see that the implementation shortfalls decrease during the training process. 

Code available at: \url{https://github.com/AI4Finance-Foundation/FinRL-Meta/tree/master/tutorials/2-Advance/execution_optimizing}

\subsection{Explainable Financial Reinforcement Learning}

\begin{figure}
\centering
\includegraphics[scale=0.55]{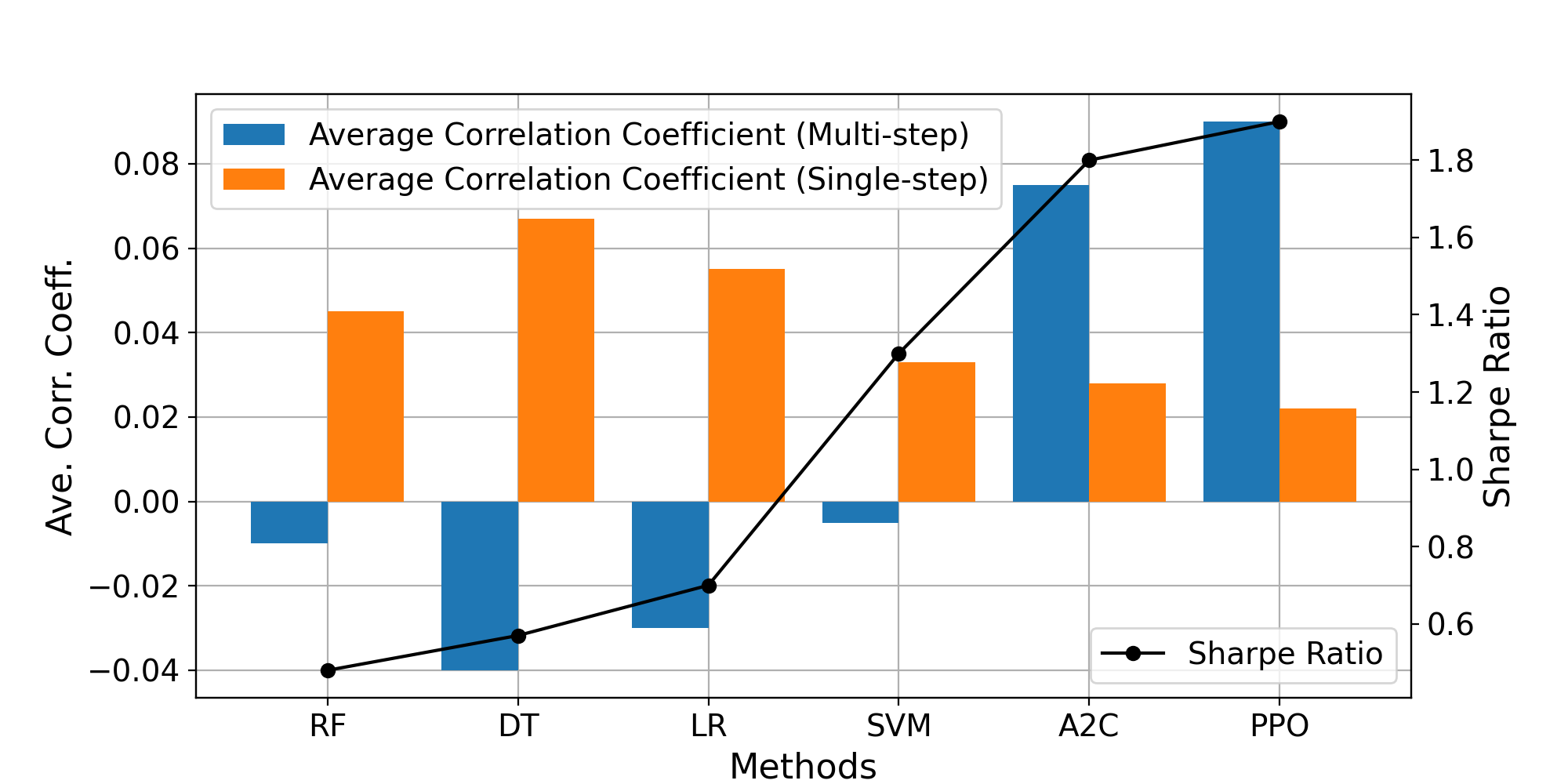}
\caption{Reproducing portfolio management of \cite{guan2021explainable}: Comparison of average correlation coefficient and Sharpe ratio among ML and DRL methods.} 
\label{fig:explainable_CumulativeReturn}
\vspace{-2mm}
\end{figure}

We reproduce \cite{guan2021explainable} that compares the performance of DRL algorithms with machine learning (ML) methods on the multi-step prediction in the portfolio allocation task. We use four technical indicators MACD, RSI, CCI, and ADX as features. Random Forest (RF), Decision Tree Regression (DT), Linear Regression (LR), and Support Vector Machine (SVM) are the ML algorithms in comparison. We use data from Dow Jones 30 constituent stocks to construct the environment. We use data from 04/01/2009 to 03/31/2020 as the training set and data from 04/01/2020 to 05/31/2022 for backtesting. In Fig.~\ref{fig:explainable_CumulativeReturn}, the results show that DRL methods have higher Sharpe ratio than ML methods. Also, DRL methods' average correlation coefficient are significantly higher than that of ML methods (multi-step).

We reproduce \cite{guan2021explainable} that compares the performance of DRL algorithms with machine learning (ML) methods on the multi-step prediction in the portfolio allocation task. We use four technical indicators MACD, RSI, CCI, and ADX as features. Random Forest (RF), Decision Tree Regression (DT), Linear Regression (LR), and Support Vector Machine (SVM) are the ML algorithms in comparison. We use data from Dow Jones 30 constituent stocks to construct the environment. We use data from 04/01/2009 to 03/31/2020 as the training set and data from 04/01/2020 to 05/31/2022 for backtesting.

Code available at: \url{https://github.com/AI4Finance-Foundation/FinRL/blob/master/tutorials/2-Advance/FinRL_PortfolioAllocation_Explainable_DRL.ipynb}

\subsection{Podracer on the Cloud}

We reproduce cloud solutions of population-based training, e.g., generational evolution \cite{finrl_podracer_2021} and tournament-based evolution \cite{liu2021podracer}. FinRL-Podracer can easily scale out to $\geq 1000$ GPUs, which features high scalability, elasticity and accessibility by following the cloud-native principle. 

If GPUs are abundant, users can take advantage of this benchmark to meet the real-time requirement of high-frequency trading tasks. Detailed instructions are provided on our website.

On an NVIDIA SuperPOD cloud, we conducted extensive experiments on stock trading, and found that it substantially outperforms competitors, such as OpenAI and RLlib \cite{finrl_podracer_2021}.

Code available at: \url{https://github.com/AI4Finance-Foundation/FinRL_Podracer}

\subsection{Ensemble Strategy}
The ensemble method combines different agents to obtain an adaptive one, which performs remarkably well in practice. We consider three component algorithms, Proximal Policy Optimization (PPO), Advantage Actor-Critic (A2C), and Deep Deterministic Policy Gradient (DDPG), which have different strengths and weaknesses. Using a rolling window, an ensemble agent automatically selects the best model for each test period. Again on the 30 constituent stocks of the DJIA index, we use data from 04/01/2009 to 06/30/2019 for training, and data from 07/01/2020 to 03/31/2022 for validation and testing through a quarterly rolling window.

Code available at: \url{https://github.com/AI4Finance-Foundation/FinRL/blob/master/tutorials/2-Advance/FinRL_Ensemble_StockTrading_ICAIF_2020.ipynb}

\subsection{Market Simulator}

\textbf{Gym market environments} \cite{brockman2016openai}

We build all of our environments following OpenAI-gym style. A first reason is this makes convenience for plugging in any of the three DRL libraries (Stable Baseline3, RLlib, ElegantRL). Another reason is user friendly. Newcomers can learn our environments faster and build their own task-specific environments.

\textbf{Synthetic data generation}:

We simulate the market from the playback of historical limit-order-book-level data and the order matching mechanism. Currently, we simulate at the minute level (i.e. one time step = one minute), which is changeable. The state is a stack of market indicators and market snapshots from the last few time steps. The action is to place an order. We support market orders and limit orders. We also provide several wrappers to accept typically discrete or continuous actions. Rewards can be configured by the participants with the aim of generating policies that optimize pre-specified indicators. In our simulator, we take into account the following factors: 1) temporary market impact; 2) order delay. We do not consider the following factors in our simulator: 1) permanent market impact of limit orders; 2) non-resiliency limit order book.

Code available at: \url{https://github.com/AI4Finance-Foundation/Market_Simulator}

\newpage

\section{Dataset Documentation and Usages}

We organize the dataset documentation according to the suggested template of \textit{datasheets for datasets} \footnote{Timnit Gebru, Jamie Morgenstern, Briana Vecchione, Jennifer Wortman Vaughan, Hanna Wal- lach, Hal Daumé Iii, and Kate Crawford. Datasheets for datasets. \textit{Communications of the ACM}, 64(12):86–92, 2021.}.

\subsection{Motivation}
\begin{itemize}
    \item \textbf{For what purpose was the dataset created?}
    
    As data is refreshing minute-to-millisecond, finance is a particularly difficult playground for deep reinforcement learning.
    
    In academia, scholars use financial big data to obtain more complex and precise understanding of markets and economics. While industries use financial big data to refine their analytical strategies and strengthen their prediction models. To serve the rapidly growing FinRL community, we creates FinRL-Meta that provides data accessing from different sources and build the data to RL environments. We aim to provide dynamical RL environments that are manageable by users.
    
    We aim to build a financial metaverse, a universe of near real-market environments, as a playground for data-driven financial machine learning.
    
    \item \textbf{Who created the dataset?}
    
    FinRL-Meta is an open-source project created by the FinRL community. Contents of FinRL-Meta are contributed by the authors of this paper and will be maintained by members of FinRL community.
    
    \item \textbf{Who funded the creation of the dataset?}
  
    AI4Finance Foundation, a non-profit open-source community that shares AI tools for finance, funded our project. 
\end{itemize}

\begin{table*}[t] \label{tab:data_sources}
    \centering
    \renewcommand{\arraystretch}{1.35}
\resizebox{1\textwidth}{!}{
\begin{tabular}{c c c c c}
\hline
\textbf{Data Source} & \textbf{Type} & \textbf{Max Frequency} & \textbf{Raw Data} & \textbf{Preprocessed Data}\\
\hline
    Alpaca &  US Stocks, ETFs &  1 min &  OHLCV &  Prices, indicators \\
    Baostock &  CN Securities &  5 min &  OHLCV &  Prices, indicators \\
    Binance &  Cryptocurrency &  1 s &  OHLCV &  Prices, indicators \\
    CCXT &  Cryptocurrency &  1 min  &  OHLCV &  Prices, indicators \\
    IEXCloud &  NMS US securities & 1 day  & OHLCV &  Prices, indicators \\
    JoinQuant &  CN Securities &  1 min  &  OHLCV &  Prices, indicators \\
    QuantConnect &  US Securities &  1 s &  OHLCV &  Prices, indicators \\
    RiceQuant &  CN Securities &  1 ms  &  OHLCV &  Prices, indicators \\
    Tushare &  CN Securities & 1 min  &  OHLCV &  Prices, indicators \\
    WRDS &  US Securities &  1 ms  &  Intraday Trades & Prices, indicators \\
    YahooFinance &  US Securities & 1 min  &  OHLCV  &  Prices, indicators \\
    AkShare &  CN Securities & 1 day  &  OHLCV &  Prices, indicators \\
    findatapy &  CN Securities & 1 day  &  OHLCV &  Prices, indicators \\
    pandas\_datareader &  US Securities &  1 day &  OHLCV & Prices, indicators \\
    pandas-finance &  US Securities &  1 day  &  OHLCV  & Prices, indicators \\
    ystockquote &  US Securities &  1 day  &  OHLCV & Prices, indicators \\
    Marketstack & 50+ countries &  1 day  &  OHLCV & Prices, indicators \\
    finnhub & US Stocks, currencies, crypto &   1 day &  OHLCV  & Prices, indicators \\
    Financial Modeling prep & US stocks, currencies, crypto &  1 min &  OHLCV  & Prices, indicators \\
    EOD Historical Data &  US stocks, and ETFs &  1 day  &  OHLCV  & Prices, indicators \\
    Alpha Vantage & Stock, ETF, forex, crypto, technical indicators &  1 min &  OHLCV  & Prices, indicators \\
    Tiingo & Stocks, crypto &  1 day  &  OHLCV  & Prices, indicators \\
    Quandl & 250+ sources &  1 day  &  OHLCV  & Prices, indicators \\
    Polygon &  US Securities &  1 day  &  OHLCV  & Prices, indicators \\
    fixer &  Exchange rate &  1 day &  Exchange rate & Exchange rate, indicators \\
    Exchangerates &  Exchange rate &  1 day  &  Exchange rate & Exchange rate, indicators \\
    Fixer &  Exchange rate &  1 day  &  Exchange rate & Exchange rate, indicators \\
    currencylayer &  Exchange rate & 1 day  &  Exchange rate & Exchange rate, indicators \\
    currencyapi &  Exchange rate & 1 day &  Exchange rate & Exchange rate, indicators \\
    Open Exchange Rates &  Exchange rate &  1 day  &  Exchange rate & Exchange rate, indicators \\
    XE &  Exchange rate &  1 day  &  Exchange rate & Exchange rate, indicators \\
    Xignite &  Exchange rate &  1 day  &  Exchange rate & Exchange rate, indicators \\

\hline
\end{tabular}}
\caption{Supported data sources. OHLCV means open, high, low, and close prices; volume data.}
\end{table*}

\subsection{Composition}
\begin{figure*}
\centering
\includegraphics[scale = 0.2]{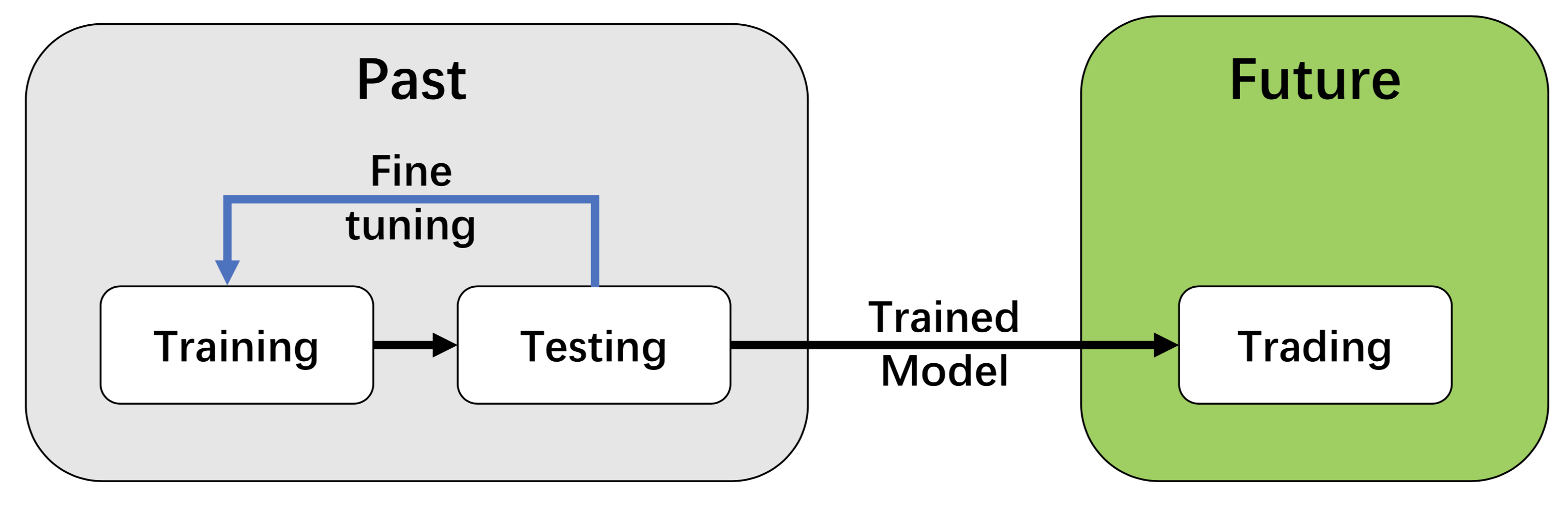}\vspace{-0.1in}
\caption{Overview of the training-testing-trading pipeline in FinRL-Meta.}
\vspace{-0.1in}
\label{fig:finrl-meta timeline}
\vspace{-2mm}
\end{figure*}

\begin{itemize}
    \item \textbf{What do the instances that comprise the dataset represent?}
    
    Instances of FinRL-Meta are financial data includes: stocks, securities, cryptocurrencies, etc. FinRL-Meta provides hundreds of market environments through an automatic pipeline that collects dynamic datasets from real-world markets and processes them into standard gym style market environments. FinRL-Meta also benchmarks popular papers as stepping stones for users to design new trading strategies.
    
    \item \textbf{How many instances are there in total?}
    
    FinRL-Meta does not store data directly. Instead, we provide codes for data accessing, data cleaning, feature engineering, and building into RL environments. Table~\ref{tab:data_sources} provides the supported data sources of FinRL-Meta. 
    
    At the moment, there are hundreds of market environments, tens of tutorials and demos, and several benchmarks provided.
    
    \item \textbf{Does the dataset contain all possible instances or is it a sample of instances from a larger set?}
    
    With our provided codes, users could fetch data from the data source by properly specifying the starting date, ending date, time granularity, asset set, attributes, etc. 
    
    \item \textbf{What data does each instance consist of?}
    
    Now there are several types of financial data, as shown in Table \ref{tab:data_sources}:
    \begin{itemize}
        \item Stocks
        \item Cryptocurrencies
        \item Securities
        \item ETFs
        \item Exchange rate:
    \end{itemize}
    
    \item \textbf{Is there a label or target associated with each instance?}
    
    No. There is not label or preset target for each instance. But users can use our benchmarks are baselines. 
    
    \item \textbf{Is any information missing from individual instances?}
    
    Yes. In several data sources, there are missing values and we provided standard preprocessing methods.
    
    \item \textbf{Are relationships between individual instances made explicit?}
    
    Yes. An instance is a sample set of the market of interest.
    
    \item \textbf{Are there recommended data splits?}
    
    We recommend users to follow our training-testing-training pipeline, as shown in Fig.~\ref{fig:finrl-meta timeline}. Users can flexibly choose their preferred settings, e.g., in stock trading task, our demo access Yahoo! Finance database and use data from 01/01/2009 to 06/30/2020 for training and data from 07/01/2020 to 05/31/2022 for backtesting.
    
    \item \textbf{Are there any errors, sources of noise, or redundancies in the dataset?}
    
    For the raw data fetched from different sources, there are noise and outliers. We provide codes to process the data and built them into standard RL gym environment.
    
    \item \textbf{Is the dataset self-contained, or does it link to or otherwise rely on external resources?}
    
    It is linked to external resources. As shown in Table~\ref{tab:data_sources}, FinRL-Meta fetch data from data sources to build gym environments.
    
    \item \textbf{Does the dataset contain data that might be considered confidential?}
    
    No. All our data are from publicly available data sources.
    
    \item \textbf{Does the dataset contain data that, if viewed directly, might be offensive, insulting, threatening, or might otherwise cause anxiety?}
    
    No. All our data are numerical.
    
\end{itemize}

\subsection{Collection Process}

\begin{itemize}
    \item \textbf{How was the data associated with each instance acquired?}
    
    FinRL-Meta fetch data from data sources. as shown in Table~\ref{tab:data_sources}.
    
    \item \textbf{What mechanisms or procedures were used to collect the data?}
    
    FinRL-Meta provides dynamic market environments that are built according to users' settings. To achieve this, we provide software APIs to fetch data from different data sources. Note that some data sources requires accounts and passwords or have limitations on number or frequency of requests.
    
    \item \textbf{If the dataset is a sample from a larger set, what was the sampling strategy?}
    
    It is dynamic, depending on users' settings, such as the starting date, ending date, time granularity, asset set, attributes, etc.
    
    \item \textbf{Who was involved in the data collection process and how were they compensated?}
    
    Our codes collect publicly available market data, which is free.
    
    
    \item \textbf{Over what timeframe was the data collected?}
    
    It is not applicable because the environments are created dynamically by running the codes to fetch data in real time.
    
    \item \textbf{Were any ethical review processes conducted?}
    
    No?
    
\end{itemize}

\subsection{Preprocessing/cleaning/labeling}

\begin{itemize}
    \item \textbf{Was any preprocessing/cleaning/labeling of the data done?}
    
    Yes. For the raw data fetched from different sources, there are noise and outliers. We provide codes to process the data and built them into standard RL gym environment.
    
    \item \textbf{Was the “raw” data saved in addition to the preprocessed/cleaned/labeled data}
    
    The raw data are hold by different data sources (data providers).
    
    \item \textbf{Is the software that was used to preprocess/clean/label the data available?}
    
    Yes. We use our own codes to do cleaning and preprocessing.
    
\end{itemize}

\subsection{Uses}

\begin{itemize}
    \item \textbf{Has the dataset been used for any tasks already?}
    
    Yes. Thousands of FinRL community members use FinRL-Meta for learning and research purpose.  Demos and tutorials are mentioned in Section \ref{benchmarks}.
    
    \item \textbf{Is there a repository that links to any or all papers or systems that use the dataset?}
    
    1. Research papers that used FinRL-Meta are list here:
    
    \url{https://github.com/AI4Finance-Foundation/FinRL/blob/master/tutorials/FinRL_papers.md}
    
    Our workshop version of FinRL-Meta \cite{finrl_meta_2021} appeared in NeurIPS 2021 Workshop on Data-Centric AI.
    
    2. The following three repositories has incorporated FinRL-meta:
    \begin{itemize}
        \item FinRL-meta corresponding to the market layer of FinRL (5.6K stars):\\ \url{https://github.com/AI4Finance-Foundation/FinRL}
        \item ElegantRL (2.1K stars) supports FinRL-Meta:\\ \url{https://github.com/AI4Finance-Foundation/ElegantRL} 
        \item FinRL-Podracer: \\ \url{https://github.com/AI4Finance-Foundation/FinRL_Podracer}
    \end{itemize}
    
    \item \textbf{What (other) tasks could the dataset be used for?}
     
     Besides the current tasks (tutorial, demo and benchmarks), FinRL-Meta will be useful for the following tasks:
     \begin{itemize}
         \item \textbf{Curriculum learning for agents}: Based on FinRL-meta (a universe of market environments, say $\geq 100$), one is able to construct a environment by sampling data samples from multiple market datasets, similar to XLand \cite{team2021open}. In this way, one can apply the curriculum learning method \cite{team2021open} to train a generally capable agent for several financial tasks.
         \item  To improve the performance for the large-scale markets, we are exploiting GPU-based massive parallel simulation such as Isaac Gym \cite{makoviychuk2021isaac}.
         \item It will be interesting to explore the evolutionary perspectives \cite{gupta2021embodied,scholl2021market,finrl_podracer_2021, liu2021podracer} to simulate the markets. We believe that FinRL-Meta will provide insights into complex market phenomena and offer guidance for financial regulations. 
     \end{itemize}
  
    \item \textbf{Is there anything about the composition of the dataset or the way it was collected and preprocessed/cleaned/labeled that might impact future uses?}
    
    We believe that FinRL-Meta will not encounter usage limit. Our data are fetched from different sources in real time when running the codes. However, there may be one or two out of $\geq 30$ data sources (in Table \ref{tab:data_sources}) change data access rules that may impact future use. So please refer to the rules and accessibility of certain data source when using.
     
    \item \textbf{Are there tasks for which the dataset should not be used?}
    
    No. Since there are no ethical problems for FinRL-Meta, users could use FinRL-Meta in any task as long as it does not violate laws.
    
    \textbf{Disclaimer: Nothing herein is financial advice, and NOT a recommendation to trade real money. Please use common sense and always first consult a professional before trading or investing.}
\end{itemize}

\subsection{Distribution}

\begin{itemize}
    \item \textbf{Will the dataset be distributed to third parties outside of the entity (e.g., company, institution, organization) on behalf of which the dataset was created?}
    
    No. It will always be held on GitHub under MIT license, for educational and research purpose.
    
    \item \textbf{How will the dataset will be distributed?}
    
    Our codes and existing environments are available on GitHub FinRL-Meta repository \url{https://github.com/AI4Finance-Foundation/FinRL-Meta}.
    
    \item \textbf{When will the dataset be distributed?}
    
    FinRL-Meta is publicly available since Feburary 14th, 2021.
    
    \item \textbf{Will the dataset be distributed under a copyright or other intellectual property (IP) license, and/or under applicable terms of use (ToU)?}
    
    FinRL-Meta is distributed under MIT License, for educational and research purpose.
    
    \item \textbf{Have any third parties imposed IP-based or other restrictions on the data associated with the instances?}
    
    No.
    
    \item \textbf{Do any export controls or other regulatory restrictions apply to the dataset or to individual instances?}
    
    No. Our data are fetched from different sources in real time. However, there may be one or two out of $\geq 30$ data sources (in Table \ref{tab:data_sources}) change data access rules that may impact future use. So please refer to the rules and accessibility of certain data source when using.
\end{itemize}

\subsection{Maintenance}

\begin{itemize}
    \item \textbf{Who will be supporting/hosting/maintaining the dataset?}
    
    FinRL-Meta has been actively maintained by FinRL community (including the authors of this paper) that has over 10K members at the moment. We are still actively updating market environments, to serve the rapidly growing FinRL community.
    
    \item \textbf{How can the owner/curator/manager of the dataset be contacted?}
    
    To contact the main developers, we encourage users join 
    our Slack channel: \\
    \url{https://join.slack.com/t/ai4financeworkspace/shared_invite/zt-v670l1jm-dzTgIT9fHZIjjrqprrY0kg} \\
    or our mailing list:\\ \url{https://groups.google.com/u/1/g/ai4finance}, \\
    
    \item \textbf{Is there an erratum?}
    
    Users can use GitHub to report issues/bugs, and use Slack channel or mailing list to discuss solutions. FinRL community is actively improving the codes, say extracting technical indicators, evaluating feature importance, quantifying the probability of backtesting overfitting, etc.
    
    \item \textbf{Will the dataset be updated?}
    
    Yes, we are actively updating FinRL-Meta's codes and data sources. Users could get information and the newly updated version through our GitHub repository, or join the mailing list: \url{https://groups.google.com/u/1/g/ai4finance}.
    
    \item \textbf{If the dataset relates to people, are there applicable limits on the retention of the data associated with the instances}
    
    The data of FinRL-Meta do not relate to people.
    
    \item \textbf{Will older versions of the dataset continue to be supported/hosted/maintained?}
    
    Yes. All versions can be found on our GitHub repository.
    
    \item \textbf{If others want to extend/augment/build on/contribute to the dataset, is there a mechanism for them to do so?}
    
    We maintain FinRL-Meta on GitHub. Users can use GitHub to report issues/bugs, and use Slack channel or mailing list to discuss solutions. We welcome community members to submit pull requests through GitHub.
    
\end{itemize}

\newpage

\section{Open Source FinRL-Meta and DAO, DeFi, NFT, Web3}

Over the past decades, we are witnessing capital growth exceeding economic growth globally. However, the door to personal capital growth is not open to all. In a way, one needs to start rich to get richer. The situation is even worsened by the competition with computers. Today in major stock markets, at least $60\%$ of the trades are automated by algorithms. 
    
How to \textit{democratize opportunity} for personal capital growth? 
We need to ally with the computers to take advantage of unprecedented amounts of data and unparalleled computing infrastructure. Therefore, we come up with this vision to establish an AI framework where retail traders can grow personal capital via a pay-by-use mode. Today, we have reached a full-stack solution that provides financially optimized deep reinforcement learning algorithms together with cloud native solutions. Users can easily access numerous financial data, as well as computational resources whenever needed.

FinRL \cite{liu2020finrl,liu2021finrl} is the first open-source framework to demonstrate the great potential of financial reinforcement learning \cite{hambly2021recent}. Over several years' development\footnote{We began to build an open-source community with a practical demonstration \cite{Xiong2018PracticalDR} at NeurIPS 2018 conference.}, it has evolved into an ecosystem, FinRL-Meta, serving as a playground for data-driven financial reinforcement learning. We believe FinRL-Meta will reshape our financial lives, while our open-source community will make sure it is for the better.

Next, we discuss the potential of combining our open-source FinRL-Meta ecosystem with the emerging technologies, such as DAO, DeFi, NFT\footnote{Ethereum: https://ethereum.org/en/}, and Web3\footnote{Web3 Foundation: https://web3.foundation/}.
\begin{itemize}
    \item Decentralized Autonomous Organizations (DAO) has three features: 1). Member-owned communities without centralized leadership; 2). A safe way to collaborate with Internet strangers; and 3). A safe place to commit funds to a specific cause. In FinRL-Meta's open-source community (over $10$K users at the moment of finalizing this paper version), we plan to encourage community members to organize into DAO funds (essentially a distributed fund) in an ad hoc manner. Members of a DAO fund will employ smart contracts to crowdfunding, design strategy, implement algorithmic trading, share profit and loss, etc.
    \item Decentralized Finance (DeFi) has three features: 1). A global, open alternative to the current financial system; 2). Products that let you borrow, save, invest, trade, and more; and 3). Based on open-source technology that anyone can program with. We would like to encourage community members to actively explore the potential of financial reinforcement learning technologies in the emerging DeFi based financial markets. For example, using the automated environment layer in Section \ref{sect:env_layer}, we will connect DeFi trading systems into gym-style market environments; also, we are actively working with financial data providers and cloud computing providers to  upgrade the trading infrastructure.
    \item Non-Fungible Token (NFT) has three features: 1). A way to represent anything unique as an Ethereum-based asset; 2). NFTs are giving more power to content creators than ever before; and 3). Powered by smart contracts on the Ethereum blockchain. In FinRL-Meta community, we would like to encourage community members to release trading strategies and codes in the NFT forms, while a DAO fund will buy these NFTs via smart contracts and then trade in a DeFi system.
    \item Third-Generation of WWW (Web3) incorporates concepts such as decentralization, blockchain technologies, and token-based economics, which is essentially community-run Internet. Our open-source FinRL-meta community will serve Web3 users a unique infrastructure and playground, where Web3 users can establish DAO funds to trade in a DeFi system.
\end{itemize}

As a near-term development, FinRL-Meta community would embrace the Massive Open Online Courses (MOOC) paradigm. For demonstrative and education purposes, our community members  are actively creating notebooks, blogs and videos. We believe it would be an effective incentive to release them in the form of NFTs. Web3MOOC guarantees authorship and ownership (a right to profit) of those NFTs, thus provides a strong incentive to our community members for content creation.

\section{Data Privacy, Strategy Privacy and Federated Learning Technology}

As our open-source community is continuously developing new features to ensure that FinRL-Meta provides better user experience, one of the main targets for our next step is to enhance financial data privacy as well as strategy privacy for users. We discuss the potential of integrating the federated learning technology into our open-source FinRL-Meta, in order to achieve data privacy and strategy privacy for our users, say collaboratively training.

Federated learning is a method for training machine learning models from distributed datasets that remains private to data owners. It allows a central machine learning model to overcome the \textit{isolated data island}, i.e., to learn from data sets distributed on multiple devices that do not  reveal or share the data to a central server. We understand that in certain scenarios, our users face the problem that they only have a small amount of financial data and are not unable to train a robust model, but they are also hesitate to train their models with others due to privacy reason. To the best of our knowledge, we would like to name a few representative examples on the use of federated learning to help the financial applications. Byrd et al.~\cite{byrd2020differentially} present a privacy-preserving federated learning protocol on a real-world credit card fraud dataset for the development of federated learning systems. The researchers in WeBank~\cite{liu2021fate} created FATE, an industrial-grade project that supports enterprises and institutions to build machine learning models collaboratively at large-scale in a distributed manner. FATE has been adopted in real-world applications in finance, health and recommender systems. We also want to mention one research work~\cite{kairouz2021advances} which points out the open problems in federated learning. We believe that, as a newly introduced method, federated learning has a lot of undiscovered and exciting applications that we can develop. We would like to encourage our community members to explore the potential of federated learning technologies for the benefit of financial applications. 

\newpage
\section{Accessibility, Accountability, Maintenance and Rights}

FinRL-Meta is an open-source project on GitHub. We use the MIT License for research and educational usage. while users can utilize them as stepping stones for customized trading strategies. Codes, market environments, benchmarks and documentations are available on the GitHub repository:\\ \url{https://github.com/AI4Finance-Foundation/FinRL-Meta}.

FinRL-Meta has been actively maintained by FinRL community that has over $10$K members at the moment. On GitHub, we keep updating our codes, merging pull requests, and fix bugs and issues. We welcome contributions from community members, researchers and quant traders.

We have accumulated six competitive advantages over the past five years. The first three are technology innovations: 
\begin{itemize}
    \item FinRL \cite{liu2020finrl,liu2021finrl} is the first framework to provide an automatic pipeline for financial reinforcement learning.
    \item For financial big data, the FinRL-meta project connects with > 30 market data sources.
    \item For cloud solutions, the FinRL-Podracer project \cite{finrl_podracer_2021} \cite{liu2021podracer} scales out to $\geq 1000$ GPUs. We have extensive testings on NVIDIA’s DGX-2 SuperPod platform.
\end{itemize}

Based on the above projects and active contributions, an open-source community in the intersection of ML and Finance fields is emerging. Our AI4Finance community is robust with the following three features:
\begin{itemize}
    \item We have over $10K$ active community members, many of which are actively designing strategies and connecting with paper trading, even live trading. We are collaborating with  tens of universities and research institutes, and $\geq 50$ software engineers from big IT companies.
    \item Both Columbia University (Department of Electrical Engineering, Department of Statistics) and New York University ((Department of Finance) have opened delicate courses about FinRL, while $\geq 120$ students in total have taken it.
    \item In academia, we have several accepted papers and also delivered several invited talks. Our AI4Finance Foundation (\url{https://github.com/AI4Finance-Foundation}) serves as a bridge between machine learning, data science, operation research, and finance communities.
\end{itemize}
As the authors of this paper and core developers of FinRL-Meta, we bear all responsibility in case of violation of rights.
\\
\\
\\

\textbf{Disclaimer: Nothing in this paper and the FinRL-Meta repository is financial advice, and NOT a recommendation to trade real money. Please use common sense and always first consult a professional before trading or investing.}